\documentclass[11pt,a4paper]{article}
\pdfoutput=1
\usepackage{jheppub}
\usepackage{amsmath,amscd}
\usepackage{amsfonts}
\usepackage{amssymb}
\usepackage{graphicx}
\usepackage{color}
\usepackage[normalem]{ulem}
\usepackage{hyperref}
\usepackage[most]{tcolorbox}
\usepackage{dsfont}
\usepackage[utf8]{inputenc}

\newcommand{\RR}{\mathbb{R}} 

\newcommand{\G}{\mathcal{G}}

\def\calb         {{\cal B}}

\def\calh         {{\cal H}}

\def\call         {{\cal L}}
\def\calm         {{\cal M}}
\def\caln         {{\cal N}}
\def\calo         {{\cal O}}

\def\calq         {{\cal Q}}

\def\cals         {{\cal S}}

\def\be{\begin{equation}}
	\def\ee{\end{equation}}
\def\bea{\begin{eqnarray}}
	\def\eea{\end{eqnarray}}

\def\a{\alpha}
\def\b{\beta}
\def\h{\eta}
\def\g{\gamma}
\def\G{\Gamma}
\def\d{\delta}
\def\e{\epsilon}
\def\D{\Delta}
\def\l{\lambda}
\def\L{\Lambda}
\def\k{\kappa}

\def\m{\mu}
\def\n{\nu}
\def\o{\omega}

\def\p{\pi}
\def\P{\Pi}
\def\r{\rho}

\def\s{\sigma}
\def\S{\Sigma}

\def\x{\xi}

\def\sF{{{ F}\!\!\!\!\hskip.8pt\hbox{\raise1pt\hbox{/}}\,}}
\def\som{{{ \omega}\!\!\!\!\hskip.8pt\hbox{\raise1pt\hbox{/}}\,}}
\def\sJ{{{\rm J}\!\!\!\!\hskip.8pt\hbox{\raise1pt\hbox{/}}\,}}

\def\pa{\partial}

\def\to{\rightarrow}
\def\nonu{\nonumber \\{}}
\def\half{{1 \over 2}}

\title{The geometry of gauged (super)conformal 
	mechanics}
\author[a]{Delaram Mirfendereski,}
\author[b]{Joris Raeymaekers,}
\author[b,c]{Canberk \c{S}anl{\i}}
\author[c,d]{and Dieter Van den Bleeken}

\affiliation[a]{EUCOS, Physics Department, Baylor University,\\ Waco, TX 76798, USA}
\affiliation[b]{CEICO, Institute of Physics of the ASCR, \\ Na Slovance 2, 182 21 Prague 8, Czech Republic}
\affiliation[c]{Physics Department, Boğaziçi University\\
	34342 Bebek / Istanbul, Turkey}
\affiliation[d]{Secondary address:\\
	Institute for Theoretical Physics, KU Leuven\\
	3001 Leuven, Belgium}

\emailAdd{d\_mirfendereski@baylor.edu}
\emailAdd{joris@fzu.cz}
\emailAdd{canberk.sanli@boun.edu.tr }
\emailAdd{dieter.van@boun.edu.tr }

\abstract{Motivated by recently explored examples, we undertake a systematic study of conformal invariance in one-dimensional sigma models where an isometry group has been gauged. Perhaps surprisingly, we uncover classes of sigma models which are only scale invariant in their ungauged form and become fully conformally invariant  only after gauging. In these cases 
the target space of the gauged sigma model satisfies a  deformation  of the well-known conformal geometry constraints.  We consider  bosonic models as well as their $\mathcal{N} = 1,2,4$ supersymmetric extensions. We solve the quantum ordering ambiguities  in implementing (super-) conformal symmetry on the physical Hilbert space. Examples of our general results are furnished by the $D(2,1;0)$-invariant  Coulomb branch quiver models relevant for black hole physics. }
\arxivnumber{}
\begin{document}
	\maketitle
	
	\section{Introduction}
	Quantum mechanical systems which possess conformal symmetry, and their supersymmetric extensions,  have appeared in  a wide variety of interesting contexts  since their initial inception in \cite{deAlfaro:1976vlx}, see \cite{Fedoruk:2011aa} for a review. One notable application is to the moduli space mechanics of extremal black holes \cite{Michelson:1999dx}  and, on a more microscopic level, as an effective description for the dynamics of their constituent D-branes \cite{Anninos:2013nra}, \cite{Mirfendereski:2020rrk}. The precise connection between these descriptions and  (near-)AdS$_2$/CFT$_1$  holography, along the lines of e.g. \cite{Sen:2008yk,Maldacena:2016upp,Bena:2018bbd,Lozano:2020sae,Lozano:2020txg}, as well as pure-Higgs states and scaling black holes \cite{Bena:2012hf, Lee:2012sc, Manschot:2012rx, Beaujard:2021fsk, Chattopadhyaya:2021rdi, Descombes:2021egc} remains an interesting open question. 
	
	{A fundamental} class of models are the  bosonic nonlinear sigma models, describing the motion of a {spinless} particle  in  a Riemannian target space $\calm$.   The requirement of conformal invariance of the action then translates into {a specific} additional   geometric structure on $\calm$, as  was worked out in \cite{Michelson:1999zf, Papadopoulos:2000ka}.
	
	Many interesting applications arise when supersymmetry is added to the mix. The geometry of supersymmetric one-dimensional sigma models was systematically investigated in \cite{Coles:1990hr, Hull:1999ng}, see  \cite{Smilga:2020nte} for {a pedagogic introduction} and a comprehensive guide to the extensive literature.  Supersymmetric mechanics can be described in terms of a zoo of multiplets that are traditionally denoted by their number of (bosonic, fermionic, auxiliary) fields. The most elegant and geometric formulations 
	use  {($\caln,\caln,0$)} multiplets without auxiliary fields{, these are sometimes also referred to as Type B and arise from dimensional reduction of 2d $(\caln,0)$ supersymmetry}. The conditions for supersymmetry can then be phrased purely in terms of geometric structures on $\calm$ and the supersymmetric ground states can be identified with cohomology classes of a suitable differential complex \cite{Witten:1982df}. {T}he geometric conditions for various amounts of superconformal symmetry {for these multiplets} were worked out in \cite{Gibbons:1997iy, Michelson:1999zf}.
	
{The ($\caln,\caln,0$) sigma models play a distinguished role since other} multiplets with auxiliary fields can often be recast in terms of {them by gauging some of their isometries, see e.g. \cite{Bellucci:2005xn, Ivanov:2011gk, Smilga:2013qy}.} This sparks the question what the  geometric  conditions  are for (super-)conformal invariance in a generic one-dimensional sigma model where  a group $\mathrm{G}$ of isometries of $\calm$ has been gauged (see  \cite{Hull:1990ms} for more details on gauged sigma models). 
	In this work we will address this question systematically for bosonic {as well as ($\caln,\caln,0$)} supersymmetric sigma models {including a Lorentz coupling to} a background electromagnetic field.

It should be remarked that, when the  group of isometries $\mathrm{G}$ acts regularly without fixed points, the gauged sigma model is classically equivalent to an ungauged sigma model on the quotient space $\calm /\mathrm{G}$. In the bosonic case, the resulting sigma model on the quotient satisfies the standard conformal geometry constraints of \cite{Michelson:1999zf,Papadopoulos:2000ka}. In the supersymmetric case however, the quotienting procedure leads to a supersymmetric sigma model that is no longer formulated  in terms of $(\caln, \caln,0)$ multiplets, as it eliminates bosonic fields but leaves the number of fermionic fields intact. For these realizations of supersymmetry the conformal invariance conditions have  not yet been worked out in the literature, and our construction leads to previously unstudied superconformal theories.

An additional motivation is that 
	working with the gauged formulation in terms of  $(\caln, \caln,0)$ multiplets on  the extended space $\calm$ has significant benefits for quantizing the model.  
	 For $\caln =2$  ($\caln =4$), the extended target space $\calm$ possesses a  (hyper-)  K\"ahler structure with torsion which is lost  in the $\calm /\mathrm{G}$ formulation. For example, using the formulation as a gauged model on $\calm$,  we will be   able to give a geometric  characterization  of ground states as certain $\mathrm{G}$-invariant cohomology classes on $\calm$.
	
The paper is organized as follows.	We obtain the conditions for conformal invariance in gauged bosonic and   $\caln=1,2,4$  supersymmetric sigma models in sections \ref{Secbos} and \ref{SecNis1}, \ref{SecNis2}, \ref{SecNis4} respectively. The corresponding (super) conformal algebras are $sl(2, \RR), osp(1|2), su(1,1|1)$ and $D(2,1;\a )$.  In each case we also solve the  operator  ordering problem in the quantum theory and  obtain a realization of these algebras on the Hilbert space.  In the $\caln = 2, 4$ cases we  give  an explicit  realization of the quantum generators as operators acting on differential forms, and characterize supersymmetric ground states as  elements of the cohomology  of a suitable differential complex. Along the way we provide more details on gauged supersymmetric sigma models and work out a target space covariant  approach for conformal invariance.
		In section \ref{Sec341} we discuss an interesting class of  examples which inspired our general analysis and arises from $\caln=4$  superconformal theories formulated in terms of (3,4,1) multiplets. As was found in \cite{Delduc:2006yp, Mirfendereski:2020rrk}, these can be reformulated geometrically in terms of type B sigma models with (4,4,0) multiplets where a certain isometry group is gauged. 
	These form nontrivial examples of our construction where the model becomes $D(2,1;0)\cong \mathfrak{psu}(1,1|2)\rtimes \mathfrak{su}(2)$ superconformally invariant only after gauging. This class of models is also physically relevant as it includes the effective description of the Coulomb branch D-brane mechanics in an AdS$_2$ scaling limit of the charges, which in turn provide an infinite set of explicit models with $D(2,1;0)$ superconformal symmetry.
	
	\section{Conformal invariance in gauged bosonic sigma models}\label{Secbos}
Conformally invariant  mechanical models \cite{deAlfaro:1976vlx} possess an $sl(2,\RR)$  algebra of conserved charges. 
	The algebra is generated by the Hamiltonian $H$, a dilatation operator $D$ and a generator of special conformal transformations $K$ satisfying the Poisson brackets
		\bea
	\{ D, H \}& =&- H \label{DHconf}\\
	\{ D, K\} &=& K \label{DKconf}\\
		\{ H,K \} &=& 2D\label{HKconf}
		\eea
	
	An interesting class of models are the  bosonic nonlinear sigma models, describing the motion of a {spinless} particle  in  a Riemannian target space.  These can be further generalized to include a potential term and/or a  Lorentz coupling to a  target space   gauge potential. The requirement of conformal invariance of the action then translates into additional and {restrictive} geometric structure {of} the target space, as  was worked out in \cite{Michelson:1999zf,Papadopoulos:2000ka}.
	
	When the target space of the sigma model possesses a group of isometries, we can consider sigma models where the corresponding symmetries are gauged.   In this section we investigate the  conditions for conformal symmetry in such gauged nonlinear sigma models. As anticipated in the Introduction, we will encounter interesting models  which become conformally invariant only after gauging, and where the  target space  satisfies a deformation of the conformal geometry constraints of  \cite{Michelson:1999zf,Papadopoulos:2000ka}. The gauging procedure amounts to imposing a set of first class constraints and in this case, in a suitable basis, the last relation in (\ref{HKconf}) holds only weakly  {(i.e. on the constraint surface only)}:
	\be
	\{ H,K \} \approx 2D\label{weak1}
	\ee
	The ungauged models are already time-translation and scale invariant, but special conformal symmetry is gained only in the gauged model.
	
Upon quantization, the symmetry generators become quantum operators which organize the Hilbert space into $sl(2,\RR)$ multiplets. We will devote special attention to the  problem of finding well-ordered quantum operators which realize the conformal algebra  on the physical Hilbert space of the gauged sigma model.
	
	\subsection{Gauged  sigma model with Lorentz coupling}
	We first review the gauging of  {one-dimensional} bosonic   sigma models with Lorentz coupling, referring to \cite{Hull:1990ms} for more details.
	We start  from a  bosonic particle  sigma model
	describing motion in a Riemannian target space of dimension $d$ with metric $G_{AB}(x)$, and we also include  a  coupling to a background vector potential $A_A(x)$. It will be useful to split the Lagrangian in a first- and second-order part in time derivatives as follows:
	\bea 
	L_B &=& L_B^{(1)} +  L_B^{(2)}\label{Lbos}\\
	L_B^{(1)} &=&  A_A \dot x^A \label{L1bos}\\
	L_B^{(2)} &=&  \half G_{AB} \dot x^A \dot x^B.\label{L2bos}
	\eea
	We assume in addition that the action is invariant under a set of  global symmetries of the form
	\be 
	\d x^A = \l^I k_I ^A.\label{deltaxgl}
	\ee
	where $ k_I^A (x)$ are vector fields in target space. 
	Closure and Jacobi identities  of the  algebra of symmetry transformations require that 
	\be 
	\call_{k_I} k_J = f_{IJ}^{\ \ K} k_K, \label{Liealg}
	\ee
	with $f_{ IJ}^{\ \ K}$ the structure constants of a Lie algebra which we will denote as $\mathfrak{g}$. We denote the dimension of $\mathfrak{g}$ by $d_\mathfrak{g}$. In the rest of this paper, we will assume that the $\mathfrak{g}$-action can be exponentiated to the action of a Lie group $\mathrm{G}$ on the target manifold, which furthermore acts without fixed points.
	
	Invariance of the action under (\ref{deltaxgl}) leads to the conditions
	\be
	i_{k_I} F = d v_I, \qquad \call_{k_I} G_{AB} = 0,\label{defvs}
	\ee
	where $F = d A$ and  $v_I (x)$  are some potentials on target space. The second requirement states that the $k_I$ are Killing vectors. From the first relation in (\ref{defvs}) one derives
	$d ( \call_{k_I} v_J ) = d (f_{IJ}^{\ \ K} v_K )$,
	and therefore there must exist constants $w_{IJ} = w_{[IJ]} $ such that
	\be
	\call_{k_I} v_J  = f_{IJ}^{\ \ K} v_K + w_{IJ}. \label{defws}
	\ee
	The $v_I$ are only determined up to constant shifts $v_I  \to v_I + c_I$,  under which   $w_{IJ}$ transforms 
	as \be w_{IJ} \to w_{IJ} + f_{IJ}^{\ \ K} c_K . \label{wshift}\ee
	These properties tell us that $w_{IJ}$ defines a Lie algebra cohomology class.

	Under (\ref{deltaxgl}),  $L^{(1)}$ transforms up to a total derivative, 
	\be \d L^{(1)} = {d \over dt} \left(\lambda^I (A_A k_I^A + v_I)\right).\ee
	We recall that,  in general, a symmetry  with parameter $\s$ under which the Lagrangian transforms as $\d_{\s} L = {d \over d t} B_{\s}$ leads to
	a  Noether charge $Q_\s$ given by
	\be \s  Q_{\s} =  {\pa L \over \pa \dot x^A} \d_{\s} x^A - B_{\s}.\label{chargeformulabos}\ee
	Applying this to the  to the symmetries (\ref{deltaxgl}) we  find the  associated  Noether charges $M_I$  to be
	\be 
	M_I = M_I^{(1)} +  M_I^{(2)}= - v_I+  G_{AB} \dot x^A k_I^B .\label{Msconfig}
	\ee

	We now want to gauge the global symmetry (\ref{deltaxgl}){, i.e.} modify the action so that it becomes invariant under transformations of the form (\ref{deltaxgl}) with  time-dependent parameters,
	\be 
	\d_\l x^A =  \l^I (t) k_I ^A.\label{deltalx}
	\ee
	For this purpose we introduce worldline gauge fields $a^I (t)$ with transformation law 
	\be 
	\d_\l a^I = \dot \l^I +  f^{\ \ \ I}_{JK} a^J \l^K.\label{deltala}
	\ee
	One checks, using the Jacobi identities in $\mathfrak{g}$, that the gauge algebra  closes and has the structure constants of $\mathfrak{g}$:
	\be 
	{}[\delta_{\l_1},\delta_{\l_2}]=\delta_{\l_3}\qquad\ \qquad\mbox{where}\quad \l_{3}^I=f_{JK}{}^I\lambda_1^J \lambda_2^K.
	\ee

	The gauging of the second order term in time derivatives $L^{(2)}$ is simply achieved by
	replacing time derivatives with covariant derivatives $D_t$:
	\be  L^{(1)} = \half G_{AB} D_t x^A D_t x^B .\ee
	Here, the covariant derivative is defined as
	\be 
	D_t x^A = \dot x^A -  a^I k_I^A,
	\ee
	and satisfies $\d_\l D_t x^A =  \l^I \pa_B k_I^A D_t  x^B$.

	The gauging of the Lorentz term $L^{(1)}$ is more subtle and requires use of the Noether procedure \cite{Hull:1990ms}. Similar to the gauging of the WZW term in two-dimensional sigma models \cite{Hull:1989jk}, the gauging is not always possible and imposes a  condition on the Lie algebra $\mathfrak{g}$. One finds that the gauged Lagrangian is\footnote{The action simplifies in the special case that $A$ defines a globally defined one-form such that $\call_{k_I} A =0$.
		In that case, one sees that the $v_I$ satisfying (\ref{defvs}) with $w_{IJ}=0$ are $v_I = - i_{k_I} A$ and the Lagrangian reduces to the naive  minimally coupled one,
		$L^{(1)} =   A_A D_t x^A$.}
	\be
	L_B^{(1)} =  A_A \dot x^A + a^I v_I.
	\ee
	Its gauge variation yields a term proportional to $w_{IJ}$ defined in (\ref{defws}) which cannot be cancelled by adding further terms to the action. Gauging therefore requires that we can make $w_{IJ}$ vanish by a transformation of the form (\ref{wshift}). In other words,   {the} $w_{IJ}$ {defined in \eqref{defws}} should be exact in Lie algebra cohomology and  {so} the second Lie algebra cohomology of $\mathfrak{g}$  {classifies the} obstructions to gauging \cite{Hull:1990ms}.  A necessary condition for gauging is thus that there exist $v^I$ that satisfy \eqref{defws} with $w_{IJ}=0$. From here onwards we will always work with that choice of $v^I$, hence imposing \eqref{gaugedbossum6} below.\\
	
	To summarize, we list the Lagrangian (L), gauge symmetries (GS), structural conditions (SC), algebra (A) and geometric conditions (GC)  for gauged nonlinear sigma models:
	\begin{tcolorbox}[ams align]
		{\rm L:} &&L_B^{(1)} &=  A_A \dot x^A + a^I v_I , & L_B^{(2)} &=\half G_{AB} D_t x^A D_t x^B\label{gaugedbossum1} \\
		&	&	&   & D_t x^A &:= \dot x^A -  a^I k_I ^A \nonu
		{\rm GS: }& & \d_\l x^A &=  \l^I k_I^A, &  \d_\l a^I &= \dot \l^I +  f^{\ \ \ I}_{JK} a^J \l^K\label{gaugedbossum2}  \\ {\rm A:} && {}[\delta_{\l_1},\delta_{\l_2}]&=\delta_{\l_3} & \l_{3}^I&=f_{JK}{}^I\lambda_1^J \lambda_2^K\label{gaugedbossum4} \\
		{\rm SC: }& &  \call_{k_I} k_J &= f_{IJ}^{\ \ K} k_K &  & \label{gaugedbossum3} \\
		{\rm GC: }& & i_{k_I} F &= d v_I ,& \call_{k_I} G_{AB} &= 0\label{gaugedbossum5} \\
		& & \call_{k_I} v_J  &= f_{ IJ}^{\ \ K} v_K  &  & \label{gaugedbossum6} 
	\end{tcolorbox}

	\subsection{Gauging, quotients and symplectic reduction}\label{Secsymplred}
	Before moving on we would like to clarify the relation between the ungauged and gauged models. We will do so both from the Lagrangian and Hamiltonian perspectives. From the Lagrangian point of view, the gauged model on $\calm$ is equivalent to an ungauged sigma model on the quotient manifold $\calm / \mathrm{G}$, while from the Hamiltonian point of view it  implements a symplectic reduction of the original phase space under the action of the  symmetry group $\mathrm{G}$. These two approaches are equivalent since $T^*\calm//\mathrm{G}=T^*(\calm/\mathrm{G})$, i.e. the reduced phase space is indeed the phase space of the reduced model.
	
	\subsubsection{Reduced sigma model on $\calm / \mathrm{G}$}\label{Secreducedmodel}
	Since the gauge potentials enter into the action quadratically and without time derivatives  we can integrate them out to obtain a classically equivalent model
	with Lagrangian
	\be 
	\tilde L_B = \tilde A_A \dot x^A + \tilde G_{AB} \dot x^A \dot x^B  - \tilde V,\label{reducedmodel}
	\ee
	where
	\be 
	\tilde A_A =  A_A + k_A^I v_I, \qquad \tilde G_{AB} = G_{AB}- k_{IA} G^{IJ} k_{JB},\qquad \tilde V = \half v_I G^{IJ} v_J.
	\ee
	Here $G_{IJ} := k_{I}^A k_{JA}$ is the matrix of inner products of the Killing vectors and $G^{IJ}$ is its inverse. We assume the latter exists, i.e. that the action of the symmetry group $\mathrm{G}$ is free.
	The Lagrangian \eqref{reducedmodel} is still invariant under the gauge transformations (\ref{deltalx}) and therefore describes particle motion on the $d-d_\mathfrak{g}$-dimensional  quotient space  $\calm / \mathrm{G}$. In particular, the $k_I$ are null directions of $\tilde G_{AB}$:
	\be 
	\tilde G_{AB} k_I^B =0.
	\ee
	 The tensor   $\tilde G_{AB}$ therefore has rank $d-d_\mathfrak{g}$ and describes the familiar\footnote{More explicitly, one could use the fact that the free group action by $\mathrm{G}$ gives $\calm$ the structure of a principal $\mathrm{G}$-bundle $\pi: \calm\rightarrow \calm/\mathrm{G}$. Introducing a local trivialization $(x^A)=(y^a,z^m)$ one finds $\tilde G_{AB}(x)\dot x^A \dot x^B=g_{ab}(y)\dot y^a \dot y^b$, with $g_{ab}$ a Riemannian metric on the base $\calm/\mathrm{G}$. See e.g. \cite{coquereaux1988riemannian} for details.}  dimensionally reduced metric on the quotient   $\calm / \mathrm{G}$. 
	 
	We note from (\ref{reducedmodel}) that the reduced Lagrangian generically has a potential term $\tilde V$  even though the original model did not. In this sense, the description as a gauged model on the larger space $\calm$ is simpler. We will see that,  especially when considering supersymmetric models, when quantizing the model  it is advantageous  use the formulation as a gauged model on $\calm$.

\subsubsection{Gauging as symplectic reduction}
	With a view towards quantizing the system it is useful to point out  
 that, from the canonical point of view, the gauging procedure implements a symplectic reduction of the phase space.
		Let us first illustrate the salient features of symplectic reduction (referring to \cite{Marsden:1974dsb} for more details) in  our bosonic sigma models. The phase space of the ungauged theory (\ref{Lbos}) is $2d$-dimensional with local coordinates $(x^A, p_A)$. The Hamiltonian is 
	\be
	H_{\rm ung} = \half  (p_A - A_A)  G^{AB}  (p_B - A_B).\label{Hung}
	\ee
	The Noether charges (\ref{Msconfig}),
	\be
	M_I =   k^A_I (p_A - A_A) - v_I
	\ee
	satisfy, thanks to (\ref{gaugedbossum3},\ref{gaugedbossum6}), the  Poisson brackets
	\be
	\{ M_I , M_J \} =-f^{\ \ K}_{IJ} M_K.
	\ee
	The $M_I$ are moment maps 
	for a Hamiltonian $\mathrm{G}$-action on the phase space whose infinitesimal version is
	\bea
	\d_\l x^A &:=& \l^I \{  x^A, M_I  \} = \l^I k_I^A  \\ \d_\l p_A &:=& \l^I \{   p_A,  M_I  \} = (-\partial_A k^B_I p_B+\call_{k_I}A_A)\lambda^I\,.\label{gaction}
	\eea
	Symplectic reduction with respect to the $\mathrm{G}$-action constructs a new  phase space which is obtained by quotienting the submanifold on which $ M_I (x,p) = 0$
	by the $\mathrm{G}$-action (\ref{gaction}). The reduced phase space 
	is sometimes denoted as
	\be  
	{T^*\calm//\mathrm{G}=M^{-1}\{0\}/\mathrm{G}}.
	\ee
	The $M_I =0$ submanifold has dimension $2d - d_\mathfrak{g}$ and the $\mathrm{G}$-quotient further reduces the dimension by another $d_\mathfrak{g}$, so symplectic reduction produces a 
	$2(d-d_\mathfrak{g})$-dimensional phase space.
	
It is now straightforward to see that the phase space of the gauged sigma model is precisely the reduced phase space  {$M^{-1} \{0\}/\mathrm{G}$}. 
	The canonical Hamiltonian of the gauged sigma model (\ref{gaugedbossum1})   is:
	\be 
		H = H_{\rm ung} + a^I M_I\label{Hcan}
	\ee
	The quickest route\footnote{Equivalently, one could choose to view the $a^I$ as dynamical variables. When doing so, there are primary first class constraints $\p_I \equiv {\d S \over \d \dot a^I} \approx0$, while $M_I \approx 0$ arise as secondary constraints. The discussion in the text then follows upon  partially fixing the gauge freedom  to get rid of the first set of constraints.} to uncovering the canonical structure of the theory is to not treat  the $a^I$ as  dynamical variables, but rather as Lagrange multipliers enforcing the first class constraints
	\be 
	M_I \approx 0.\label{constraint}
	\ee
	These generate the gauge transformations (\ref{gaction}) on the dynamical variables $(x^A, p_A)$, while the transformation law  (\ref{gaugedbossum2})  for the $a^I$ is the standard one for Lagrange multipliers, see \cite{Henneaux:1992ig}. 

	Since we are free to redefine the Hamiltonian by combinations of the first class constraints, we can just as well work with
	\be 
	H'= H - a^I M_I \label{Hpdef}
	\ee
which  coincides with the Hamiltonian in the ungauged model (\ref{Hung}).  It is then clear that the
	constraints (\ref{constraint}) and gauge symmetry (\ref{gaction})  implement the symplectic reduction  to  {$ M^{-1} \{0\}/\mathrm{G}$}.

	\subsection{Conditions for conformal invariance}\label{Secbosconf}
	We will now derive the further geometric conditions on the target space in order for the gauged sigma model (\ref{gaugedbossum1}) to be conformally invariant. As usual \cite{deAlfaro:1976vlx}, conformal transformations arise from a $PSL(2,\RR)$ subgroup of time reparametrizations
	\be
	t'=\frac{at+b}{ct+d}, \qquad \d t = - P (t),
	\ee
	where $P(t)$ is  a time-dependent parameter
	\be P=u+ vt+w t^2.\ee Here the constant parameters $u,v,w$ are associated to time translations, time rescalings and special conformal transformations respectively. 
	The transformation of the coordinates $x^A$ can be written covariantly \cite{Papadopoulos:2000ka} in terms of a target space vector  $\x^A$ as
	\be
	\delta_P x^A =P\dot x^A+\dot P\xi^A.\label{deltaPx}
	\ee
	For example,  if  the  $x^A$ transform as  `primary fields of dimension $\D${'},
	\be 
	x'(t')=(ct+d)^{2\D} x(t), 
	\ee
	we obtain an infinitesimal transformation of the form  (\ref{deltaPx}) with $\x^A = \D x^A$.
	For the conformal transformation of the gauge fields $a^I$ we make the ansatz 
	\be
	\delta_P a^I=P\dot a^I+\dot P(\delta^I_J+\gamma^I{}_J)a^J+\ddot P h^I,\label{deltaPa}
	\ee
	with $\gamma^I{}_J$  {constants} and $h^I$  functions of $x^A${. The potentials $h^I$ play a crucial role, in that they parameterize the deformation of the special conformal transformation of the gauge-covariant velocities. That is, one can compare
		\begin{eqnarray}
			\delta \dot x^A&=&\ddot x^A P+(\delta^A_B+\partial_B\xi^A)\dot x^B \dot P+\xi^A\ddot P\\
			\delta D_t x^A&=&\frac{d}{dt}( D_t x^A) P+(\delta^A_B+\partial_B\xi^A)D_t x^B \dot P+\xi_\perp^A\ddot P,\label{confcovder}
		\end{eqnarray}
		where $\xi_\perp$ is a target space vector defined as
		\begin{equation}
			\xi_\perp^A:=\xi^A-h^Ik_I^A.\label{xiperpdef}
		\end{equation}
		The fact that it is the the vector $\xi_\perp$, rather than $\xi$, which   appears in the last term of (\ref{confcovder}) will turn out to have important consequences for the conformal invariance conditions in gauged sigma models.
	
	Before examining invariance of the action it is useful to work out the constraints imposed by  closure of the combined algebra of gauge transformations (\ref{gaugedbossum4}) and conformal transformations
	(\ref{deltaPx},\ref{deltaPa}). This leads to  {the conditions}
	\be 
	\call_{\xi} k_I^A=-\gamma^J{}_I k_J^A, \qquad
	\call_{k_J} h^I=\gamma^I{}_J-f_{JK}{}^I h^K, \qquad
	\call_\xi h^I =\gamma^I{}_J h^J.\label{closureh}
	\ee
	One then finds the algebra
	\bea
	{}[\delta_{P_1},\delta_{P_2}]&=&\delta_{P_3}\qquad\qquad\mbox{where}\quad P_3=\dot P_1P_2-P_1\dot P_2\\
	{}[\delta_{P},\delta_{\l_1}]&=&\delta_{\l_2}\qquad\ \qquad\mbox{where}\quad
	\l_{2}^I=-P\dot \l_1-\dot P\gamma^I{}_{J}\lambda^J_1.
	\eea
	
	Now we turn to the conditions imposed by  invariance of the action  under conformal transformations (\ref{deltaPx},\ref{deltaPa}). The first order part $L^{(1)}$ is invariant provided that 
	\be 
	i_\x F = d ( h^I v_I ), \qquad {\call_\x v_I =- \g^J_{\ I} v_J}
	\ee
	and then transforms by a total derivative  
	\be 
	\d_P L^{(1)} = {d \over dt} \left( P L^{(1)} + \dot P ( i_\x A + h^I v_I ) \right).\label{Bconfbos1}
	\ee
	
	Demanding invariance of the second order Lagrangian $L^{(2)}$  fixes the last term in the transformation (\ref{deltaPa}) of $a^I$:
	\be
	h^I = G^{IJ} \x_A k_J^A\label{hsol}\,.
	\ee
	One checks that $h^I$ in (\ref{hsol}) indeed satisfies (\ref{closureh}). 
	In addition, conformal invariance imposes the following conditions on the background
	\bea
	\call_\x G_{AB} &=& - G_{AB}\label{xiCKV}\\
	\x_{\perp\, A} &=& - \half \pa_A K\label{xiperpexact}
	\eea
	The first condition states that $\x$ must be a conformal Killing vector, while the second condition requires (the one-form dual to)   $\xi_\perp$  to be exact. The definition \eqref{xiperpdef} together with the invariance condition \eqref{hsol} implies that $\xi_\perp$ is}  the projection of $\x$ orthogonal the Killing vectors $k_I$ :
	\be
	\x_\perp^A  = P_{\perp \ B}^{\ A} \x^B  \qquad  P_{\perp \ B}^{\ A} := \d^A_B - G^{IJ} k_{IB} k_J^A. 
	\ee
	The right hand side of (\ref{xiperpexact}) involves a target space  function $K(x)$, which will turn out to play the role of the special conformal Noether charge. We can actually give an explicit expression for $K$: using (\ref{gaugedbossum5},\ref{closureh},\ref{xiCKV}) and the relation $\call_{\x_\perp} h^I =0$ which follows from them, one shows that
	\be 
	K = 2 \x_{\perp A} \x_\perp^A \label{Ksol}
	\ee
	satisfies (\ref{xiperpexact}).  We note that (\ref{Ksol}) reduces to the standard expression of \cite{Michelson:1999zf} in the ungauged case.

	Let us comment more on the condition (\ref{xiperpexact}), which is required for invariance under special conformal transformations.  This condition forms the main generalization brought about by the  gauging procedure: in ungauged sigma models, conformal invariance  requires the  dual of the conformal Killing vector $\xi$ to be exact, while in the gauged case  it is sufficient that this holds for it's projection $\x_\perp$ orthogonal to the symmetry orbits.  These two conditions are different when  the conformal Killing vector  is not orthogonal to the symmetry orbits, which from (\ref{hsol}) is equivalent to the $h^I$ being nonzero. 
	Therefore, for nonvanishing $h^I$  the target space $\calm$ of the gauged sigma model does not satisfy the standard geometric constraints of \cite{Michelson:1999zf, Papadopoulos:2000ka} and the model {\em is conformally invariant only when gauged.}
	In particular, the ungauged model on $\calm$ would in this case be scale invariant but not invariant under special conformal transformations. We will discuss explicit examples where $h^I \neq 0$ in Section \ref{Sec341}. 
	
		If the  conditions (\ref{xiCKV},\ref{xiperpexact}) are met, $L^{(2)}$ transforms by a total derivative,
	\be 
	\d_P L^{(2)} = {d \over dt} \left(P L^{(2)} - \half \ddot P K \right).\label{Bconfbos2}
	\ee

	Summarizing, conformal symmetry (CS) leads to the following  structures in addition to (\ref{gaugedbossum1}-\ref{gaugedbossum6}):
	\begin{tcolorbox}[ams align]
		{\rm CS: }& &\delta_P x^A &=P\dot x^A+\dot P\xi^A & \d_P a^I &= P \dot a^I +\dot P (\g^I_{\ J} a^J + a^I) +\ddot P  h^I\label{confbossum1}\\
		{\rm A:}&&  {}[\delta_{P_1},\delta_{P_2}]&=\delta_{P_3}& P_3&=\dot P_1P_2-P_1\dot P_2\label{confbossum3}\\
		&&	{}[\delta_{P},\delta_{\l_1}]&=\delta_{\l_2} &
		\l_{2}^I&=-P\dot \l_1-\dot P\gamma^I{}_{J}\lambda^J_1\label{confbossum4}\\
		{\rm SC: }& &    \call_{\xi} k_I^A&=-\gamma^J{}_I k_J^A  & \gamma^I{}_L f_{JK}{}^{L}&=f_{LK}{}^I\gamma^{L}{}_J+f_{JL}{}^I\gamma^{L}{}_K \label{confbossum2}\\
		&&	 \call_{k_J} h^I&=\gamma^I{}_J-f_{JK}{}^I h^K, &
		\call_\xi h^I &=\gamma^I{}_J h^J \label{confbossum2b}\\
		{\rm GC: }& & i_{\x} F &= d ( h^I v_I) & \call_{\x} G_{AB} &= -G_{AB}\label{confbossum5}\\
		&& {\call_\x v_I} &=-{\g^J_{\ I} v_J}&    	h^I &= G^{IJ} \x_A k_J^A  
		\label{confbossum6}\\
		&& &&  \x_{\perp A} &:= \x_A - h^I k_{IA}= - \half \pa_A K . 	\label{confbossum7}
	\end{tcolorbox}
	
	A remark is in order before moving on to the canonical formalism and quantization. As we saw in section (\ref{Secreducedmodel}), we can equivalently describe the gauged model as a sigma model (with potential term) on the quotient space $\calm / \mathrm{G}$. Consistency requires that, in this description, the  reduced background gauge field $\tilde A_A$, metric $\tilde G_{AB}$ and potential $\tilde V$ (cfr. (\ref{reducedmodel})) do satisfy the general constraints for conformal invariance derived in \cite{Michelson:1999zf, Papadopoulos:2000ka}. One checks that this is indeed the case, since the conditions listed above imply
		\bea
		i_{\xi_\perp} \tilde F &=&0\\
		\call_{\xi_\perp} \tilde G_{AB} &=& -\tilde G_{AB}, \qquad \xi_{\perp A} = -\half \pa_A K\\
		\call_{\xi_\perp} \tilde V &=& \tilde V.
		\eea

	\subsection{Canonical formalism}
	With a view towards quantizing the system, we now want to
	work out how the conformal symmetries are represented in canonical variables. As explained in Section \ref{Secsymplred}, the phase space variables are $(x^A, p_A)$ and obey the canonical Poisson brackets
	\be 
	\{ x^A , p_B \} = \d^A_B.
	\ee
	The effect of the gauging procedure is to subject the system to the first class constraints
	\be 
	M_I =   k^A_I (p_A - A_A) - v_I \approx 0.\label{QIclbos}
	\ee 
	
	From the transformations (\ref{confbossum1}) and variations of the Lagrangian (\ref{Bconfbos1},\ref{Bconfbos2}), we compute the  conformal Noether charges. These are time-dependent, since  the field variations and  boundary terms in the variation of the Lagrangian are time-dependent,  reflecting the fact that the dilatation and special conformal generators don't commute with the Hamiltonian. Conservation means in this case that the Noether charges satisfy 
	\be 
	{\pa Q \over \pa t} + \{Q, H\}=0.
	\ee
The conformal charges evaluated at $t=0$  are
	\bea
	H' &=& \half (p_A - A_A) G^{AB} (p_B - A_B)\label{Hclbos}\\
	D 
	 &=& \x^a (p_a - A_a)  -h^I v_I \label{Dclbos} \\
	K &=& 2 \x_{\perp A} \x_\perp^A.\label{Kclbos}
	\eea
	We recall from (\ref{Hpdef}) that $H'$ differs from the canonical Hamiltonian by a combination of the constraints.
	The Poisson brackets between the conformal charges and the constraints are
	\bea
	\{ M_I, H \} &=&0\label{Hinv}\\
	\{ M_I, D \} &=& - \g^J{}_{I} M_J\label{QDPB}\\
	\{ M_I, K \} &=&0\label{Kinv}
	\eea 
	In other words, the conformal charges weakly Poisson-commute with the constraints,
	\be
	\{ M_I, H' \} \approx \{ M_I, D  \} \approx \{ M_I, K \} \approx 0.
	\ee

	Let us now compute the Poisson brackets between the conformal charges. We find
	\bea
	\{ D , H'\}& =&- H' \label{HDPB}\\
		\{ D, K\} &=& K \label{DKPB}\\
	\{ H',K \} &=& 2(D- h^I M_I ). \label{HKPB}
\eea
		The  Poisson bracket  (\ref{HKPB}) shows  that, when the $h^I$ are non-vanishing, the conformal algebra is satisfied  weakly but not strongly, as announced in (\ref{weak1}). 
		Therefore in this case the model 		{\em becomes conformally invariant only after symplectic reduction to the $M_I \approx 0$ surface.} This is just the phase-space equivalent of our findings in the Lagrangian language in Section \ref{Secbosconf}. 
	
Before discussing the quantum theory, we would like to comment	on the apparent  contrast of (\ref{HKPB})  with the algebra of transformations (\ref{confbossum3}), where conformal transformations  closed among themselves without the need for an  additional gauge transformation. 
	The reason for this is that  the time translations in (\ref{confbossum3}) are generated by the canonical Hamiltonian $H$ in (\ref{Hcan}), which differs from $H'$ by a combination of the gauge generators. The transformations generated by $H', D, K$ are therefore a combination of the earlier conformal and  gauge tranformations,
	\be 
	\d'_P := \d_P +  \d_{\l}, \qquad {\rm where\ } \l^I = - u a^I.
	\ee
	One checks that these satisfy the algebra
	\be 
	[ \d'_{P_1}, \d'_{P_2} ] = \d'_{P_3} + \d_\l,\qquad  P_3 = \dot P_1 P_2 - \dot P_2 P_1,\qquad  \l^I = 2 (u_1 w_2 - u_2 w_1) (  a^I t - h^I),
	\ee
	in agreement with (\ref{HKPB}).

	\subsection{Quantization}\label{Secbosquant}
	The quantization of the gauged one-dimensional sigma model is well studied. If the gauge group $G$ is compact then there exist two equivalent approaches \cite{tuynman1990reduction, henneaux2020quantization}: (i) as conventional quantization of motion on the quotient $\calm/G$ or (ii) by first quantizing the motion on $\calm$ and then restricting the Hilbert space to a subspace which is the common kernel of the quantum generators of $G$. In this paper we will consider option (ii) and work in a pedestrian approach to Dirac's quantization method of first-class constraints. We refer the reader to e.g. \cite{tuynman1990reduction, henneaux2020quantization} for a more detailed but also more technical discussion. Some of the subtleties we will encounter can be treated in  a more streamlined manner in the BRST formalism.
	
	In the spirit of (ii) we first construct operators\footnote{We will not make a notational distinction between classical observables and quantum operators in this work, hoping that it is clear from the context which is meant. There will be however some operator ordering ambiguities whose resolution leads to interesting quantum corrections. We'll indicate the final well-ordered operators with a prime, to distinguish them from their more naive, Hermitian but incorrect counterparts.}  $H,K,D,M_I$ acting on a Hilbert space  {$\calh_{\mathrm{large}}$}  {so that they} realize the operator version of the Poisson bracket relations (\ref{Hinv}-\ref{Kinv}) and (\ref{HDPB}-\ref{DKPB}).	The second step is to  perform  {the quantum analog of the symplectic reduction by} the constraints $M_I \approx 0${: The $M_I$ (or more precisely their well-ordered version $M'_I$, see below) are now operators on $\calh_\mathrm{large}$, and one defines the physical Hilbert space as \be \calh_{\mathrm{phys}}=\cap_I\ker M_I'\subset \calh_{\mathrm{large}}.\ee 
	 In addition, one has to specify how the inner product on $\calh_{\rm phys}$ is obtained from the one on $\calh_{\rm large}$.
	
	The first step in this procedure  is essentially  a matter of finding the correct operator ordering.
	We take our Hilbert space $\calh_{\mathrm{large}}$ to consist of square integrable functions on the target space $\calm$ with respect to the covariant inner product
	\be 
	( f_1, f_2) = \int d^d x \sqrt{ G} f_1^* f_2.\label{L2norm}
	\ee
	As usual we  represent $x^A$ as a multiplication operator and the  momentum operators and their Hermitian conjugates are\footnote{{We choose to follow the conventions of \cite{Michelson:1999zf}, alternatively one can work with Hermitian momenta $p_A^{\mathrm{H}}=(p_A^{\mathrm{H}})^\dagger=G^{-1/4}p_A G^{1/4}$ as is done for example in the classical reference \cite{DeWitt:1952js}.}}
	\be 
	p_A = - i \pa_A , \qquad p_A^\dagger = {1 \over \sqrt{G}} p_A \sqrt{G}= p_A - i \G_{AB}^B\label{momquant} 
	\ee
	Generalizing the discussion of \cite{Michelson:1999zf} to our gauged models, we start from the Noether charges (\ref{Hclbos}-\ref{Kclbos}) and  find the following candidate quantum operators 
	\begin{eqnarray}
	H' &=&\half (p_A^\dagger - A_A) G^{AB} (p_B - A_B)\label{Hna}\\
	D &=& \half  \left(\x^A p_A+p_A^\dagger \x^A\right) -\xi^A A_A - h^I v_I \\
	K &=&2 \x_\perp^A \x_{\perp A}\\
	M_I &=&  k_I^A (p_A - A_A)- v_I.\label{Mna}
	\end{eqnarray}
	One can verify that these operators are Hermitean and furthermore all restrict to operators\footnote{Any operator $\calo$ on $\calh_\mathrm{large}$ is well defined on $\calh_\mathrm{phys}\subset \calh_\mathrm{large}$, but its image when restricted to $\calh_\mathrm{phys}$ should again fall into $\calh_\mathrm{phys}$ for it to be an operator on $\calh_\mathrm{phys}$. That condition is equivalent to $[M_I,\calo]=C_I{}^J M_J$ for some operators $C_I{}^J$.} on $\cap_I\ker M_I$ since
	\begin{equation}
		[ M_I, M_J ] = - i f_{IJ}^{\ \ K} M_K, \quad\ \ 		[ M_I, H' ] = 0, \quad\ \ 
		[ M_I, K ] = 0 ,\quad\ \
		[ M_I, D ] = -i  \g^J{}_I M_J 
	\end{equation}
	A further calculation reveals that they close almost, but {\it not completely}, into an SL(2,$\mathbb{R}$) algebra on $\cap_I\ker M_I$:
	\begin{equation}
	[D,H'] = -i H', \qquad
	[D, K] = i K, \qquad 
	[H',K] = 2 i  D - 2i  h^I M_I -(\gamma^I{}_I+h^I f_{IJ}{}^J)\label{HKcommqu}
	\end{equation}
	The problem is the rightmost term in the $[H',K]$, which originates from $[M_I,h^J]=-i\call_{k_I} h^J=-i(\gamma^J{}_I-f_{IK}{}^Jh^K)$. This troublesome term can however be removed by a slight modification of the operators $D$ and $M_I$. If instead of (\ref{Hna}-\ref{Mna})  one works with
		\begin{tcolorbox}[ams align]
		H' &=\half (p_A^\dagger - A_A) G^{AB} (p_B - A_B)\label{Hbos}\\
		D' 
		 &=\half  \x_\perp^A ( p_A - A_A) + {\rm h.c.}\label{Dbos}\\
		K &=2 \x_\perp^A \x_{\perp A}\label{Kbos}\\
		M_I' &=  k_I^A (p_A - A_A)- v_I-\frac{i}{2}f_{IJ}{}^J\label{QIbos}
	\end{tcolorbox}
	then the commutation relations (\ref{HKcommqu}) get modified to\footnote{To obtain these results one has to use $f_{IJ}{}^{K}f_{KL}{}^L=0$ and $\gamma^I{}_J f_{IK}{}^K=0$ which follow from the Jacobi identity and \eqref{confbossum2} respectively.}
	\begin{align}
	[ M_I', M_J' ] = - i f_{IJ}^{\ \ K} M_K', \quad\ \ 		[ M_I', H' ] = 0, \quad\ \ 
	[ M_I', K ] = 0 ,\quad\ \
	[ M_I', D ] = 0\\
	[D',H']=-iH'+C^I M'_I\qquad [D',K']=i K'\qquad [H',K']=2i D'\label{HKcommqumod}
	\end{align}where
	\begin{equation}
	C^I=[h^I,H']=\frac{i}{2}(G^{AB}\partial_B h^I p_A+p_A^\dagger G^{AB}\partial_B h^I ).
	\end{equation}
	So one can conclude that $H', D'$ and $K$ are Hermitian operators with respect to (\ref{L2norm}) which  form an SL(2,$\mathbb{R}$) algebra on  the physical Hilbert space defined as $\calh_\mathrm{phys}=\cap_I \ker M_I'$.
	We note that the new dilatation operator $D'$ is related to the original one as
	\bea 
	D'&=&D-\frac{1}{2}(h^I M_I+M_I h^I)\\
	&=&D + \frac{i}{2}\gamma^I{}_I-h^I M'_I
	\eea
The inner product on the physical Hilbert space takes the general form \cite{henneaux2020quantization}
\be 
\langle f | g \rangle = \int d^d x \sqrt{ G} \prod_I \d (\chi_I ) \det  ( \{  \chi_J , M'_K \}) f^* g,\label{L2red}
\ee
where $\chi_I (x)$	are a suitable set of gauge-fixing functions. This inner product does not depend on the choice of $\chi_I (x)$.

To conclude let us  comment on the meaning of the expression   (\ref{QIbos}) for the quantum gauge charges $M_I'$. Their action on a wavefunction $\psi (x)$ is
\be 
i M_I '\psi = k_I^A \pa_ A \psi + \half f_{IJ}^{\ \ J}\psi -i \left( i_{k_I} A + v_I \right) \psi \label{deltagpsi}
\ee
This expression shows that the wavefunctions are not simply scalars (whose transformation would consist of only the first term), but sections of a certain bundle. More precisely, the second and last terms mean that the wavefunctions transform as densities of weight $- \half$ under redefinitions of the Killing vectors $k_I$, and with unit charge  under gauge transformations of the field $A$.
By this we mean that under
\bea
x \to x'(x)\\
k_I \to N_I^{ \ J} k_J \label{redefk}\\
A \to A + d\L
\eea
the wavefunction transforms as
\be 
\psi (x) \to \left|\det  N_I^{ \ J}\right|^{-\half} e^{ i \L } \psi \left( x (x') \right).
\ee
Recalling  that 
\be 
\call_{k_I} k_J = f_{IJ}^{\ \ k} k_K, \qquad \call_{k_I} A = d \left( i_{k_I} A + v_I \right),
\ee 
the diffeomorphism generated by $k_I$ acts on wavefunctions precisely as in (\ref{deltagpsi}). The fact that the wavefunction is a weight $-\half$ density under (\ref{redefk}) is in fact required for consistency of the reduced inner product (\ref{L2red}), since the measure transforms  as a weight one density.
 It leads to the   shift  by $\frac{i}{2}f_{IJ}{}^J$ in $M_I'$, which is a generic feature for non-unimodular\footnote{A Lie group is said to be unimodular when it admits a bi-invariant measure, this is equivalent to $f_{IJ}{}^J=0$ and examples are abelian and semi-simple Lie groups.} gauge groups $G$ 
	\cite{tuynman1990reduction}. 
	In the BRST formalism it can be  traced back to an ordering ambiguity of the ghosts in the BRST operator which gets resolved by demanding that operator to be nilpotent and Hermitian \cite{Henneaux:1992ig}. The fact that  for non-unimodular gauge groups $M'_I$ is non-Hermitian with respect to the inner product in the large space (\ref{L2norm}) does not present any physical problem.}

	\section{Gauged sigma models with $ osp (1|2)$ superconformal symmetry}\label{SecNis1}
	In this and the following sections we extend the analysis of conformal invariance to gauged sigma models  including fermions and possessing various amounts of supersymmetry. We focus exclusively on models with ($\caln,\caln,0$) multiplets, i.e. type B supersymmetry which arises from dimensional reduction of 2-dimensional sigma models with chiral fermions and supersymmetry.
	
	\subsection{The gauged  $\caln = 1B$ supersymmetric sigma model}
	Quantum mechanical sigma models involving $d$ bosonic fields $x^A$ and possessing $\caln = 1B$ Poincar\'{e}  supersymmetry include  $d$ real fermionic superpartners $\chi^A = (\chi^A)^*$. Supersymmetry transformations act as
	\bea 
	\d_\e x^A &=& - i \e \chi^A \label{Nis1Bglx}\\
	\d_\e \chi^A &=& \e \dot x^B,\label{Nis1Bgl}
	\eea 
	where $\e = \e^*$ is a real fermionic parameter\footnote{ Note that, in our conventions, complex conjugation reverses the order of the fermions so that the right-hand side of (\ref{Nis1Bglx}) is real.}.
	The $\caln = 1B$ supersymmetric completion of the bosonic Lagrangians $L^{(1)}$  in (\ref{L1bos})  and  $L^{(2)}$  in (\ref{L2bos})  is obtained by adding the following fermionic parts \cite{Michelson:1999zf}:
	\bea 
	L_F^{(1)} &=&  - {i \over 2} F_{AB} \chi^A \chi^B  \label{L1ferm}\\
	L_F^{(2)} &=&  {i \over 2}  G_{AB}\chi^A \check \nabla_t \chi^B -  {1 \over 12} \pa_{[A} C_{BCD]}  \chi^A  \chi^B  \chi^C  \chi^D ,\label{L2ferm}
	\eea
	where 
	\bea 
	\check \nabla_t \chi^a &:=& \dot \chi^A + \check \G^A_{\ BC} \dot x^B \chi^C\\
	\check \G^A_{\ BC} &:=& \G^A_{\ BC} + \half  C^A_{\ BC}. \label{torsionder}
	\eea
	We note that the general $L_F^{(2)} $ includes a  target space three-form  $C_{ABC} = C_{[ABC]}$ which is not restricted by $\caln = 1B$ supersymmetry.  The $\check \G^A_{\ BC}$ in (\ref{torsionder}) are the {connection} coefficients of a torsionful covariant derivative $\check \nabla$  acting on target space tensors as
	\be
		\check \nabla_A v^B_C := \pa_A v^B_C + \check \G^B_{ AD} v^D_C - \check \G^D_{ AC} v_D^B. \label{checknabladef}
		\ee 

	We now want to find the gauged version of the above $\caln = 1B$ sigma models. The gauging of the bosonic part led to (\ref{gaugedbossum1}-\ref{gaugedbossum6}) and we now focus on the gauging    of the fermionic Lagrangians $L^{(1)}_F$ and $L^{(2)}_F$ assuming the gauge transformations and geometric properties listed in (\ref{gaugedbossum1}-\ref{gaugedbossum6}).
	The first order Lagrangian $L^{(1)}_F$ is gauge invariant provided that the fermions $\chi^A$ transform in the same way as $D_t x^A$, i.e.
	\be 
	\d_\l \chi^A = \l^I \pa_B k^A_I \chi^B .\label{gaugechi}
	\ee
	The second order Lagrangian $L^{(2)}_F$ is invariant under global symmetries with constant $\l^I$  provided that the torsion  is an  invariant tensor\footnote{One needs the identity that, for $k$ a Killing vector, 
		$ \call_{k} \G^C_{\ AB} =- \pa_A \pa_B k^C  $.},
	\be 
	\call_{k_I } C_{ABC}  = 0.
	\ee
	To gauge $L^{(2)}_F$ it then suffices to replace $\check \nabla_t \chi^A$ by a gauge-covariant version $\check D_t \chi^A$ satisfying
	\be 
	\d_\l \check D_t \chi^A = \l^I \pa_C k^A_I \check D_t \chi^C .
	\ee
	One finds that the latter requirement determines
	\be  
	\check D_t \chi^A := \check \nabla_t \chi^A +  a^I \left( \nabla^A k_{IB} + \half C^A_{\  BC} k_I^C\right) \chi^B.
	\ee
	
	One  verifies that the total gauged action  is $\caln = 1B$ supersymmetric provided the supersymmetry transformations (\ref{Nis1Bgl}) are gauge-covariantized and the gauge fields $a^I$ are  singlets under supersymmetry:
	\bea 
	\d_\e x^A &=& - i \e \chi^A\\
	\d_\e \chi^A &=& \e D_t x^A\\
	\d_\e a^I &=&0.\label{Nis1Bgauged}
	\eea 
	The total Lagrangian transforms by a total derivative:
	\be
	\d_\e L_{\rm tot} = \e {d \over dt}\left( - i  A_A \chi^A - {i \over 2} G_{AB} D_t x^A \chi^B-{1\over 12} C_{ABC} \chi^A \chi^B \chi^C  \right).\label{Bgauge}
	\ee

	To summarize, the $\caln = 1B$ supersymmetric extensions of the gauged sigma   models   (\ref{gaugedbossum1}) have the following structures in the fermionic sector:
	\begin{tcolorbox}[ams align]
		{\rm L:}&&	L_F^{(1)} &= - {i \over 2} F_{AB} \chi^A \chi^B  , & L_F^{(2)} &= {i \over 2}  G_{AB}\chi^A \check D_t \chi^B -  {1 \over 12} \pa_{[A} C_{BCD]}  \chi^A  \chi^B  \chi^C  \chi^D \label{gaugedfermsum1} \\
		&&	&   & \check D_t \chi^A &:= \check \nabla_t \chi^A +  a^I \left( \nabla^A k_{IB} + \half C^A_{\  BC} k_I^C\right) \chi^B \nonu
		&&	&   & \check \nabla_t \chi^A &:= \dot \chi^A + \left( \G^A_{\ BC} + \half  C^A_{\ BC} \right) \dot x^B \chi^C\nonu
		{\rm GT:}&& \d_\l \chi^A &= \l^I \pa_B k^A_I \chi^B  & &\label{gaugedfermsum2}  \\
		{\rm GC:}&& && \call_{k_I}C_{ABC} &= 0
		\label{gaugedfermsum3} 
	\end{tcolorbox}


	\subsection{Conditions for $osp(1|2)$ superconformal invariance}
	We now investigate the additional conditions for conformal invariance of the gauged supersymmetric sigma model, assuming that the invariance conditions for the bosonic part (\ref{confbossum1}-\ref{confbossum6}) are met.  We will see that the torsion tensor
	$C_{ABC}$ needs to satisfy additional constraints.

	First we need to specify the conformal transformation law for the fermions. We postulate the transformation law\footnote{ One way to derive this  is by demanding that the first order Lagrangian $L^{(1)}_F$  is conformally invariant without additional conditions on the background. }  
	\be 
	\d_P\chi^A = P \dot \chi^A + \dot P \left( \pa_B \x^A \chi^B + \half \chi^A \right).
	\ee
	Invariance of the second order term $L^{(2)}_F$ leads to the following conditions on the torsion tensor:
	\bea
	\call_\x C_{ABC} &=& - C_{ABC}\label{Ccond1}\\
	C_{ABC}\x_{\perp}^C &=& 2\pa_{[A} h^I k_{I B]}.\label{Ccond2}
	\eea
	These generalize the conditions on the torsion in the ungauged case (see (3.12) in \cite{Michelson:1999zf}).  
	If (\ref{Ccond1},\ref{Ccond2}) are satisfied, the variation of the total fermionic Lagrangian $L_F =  L_F^{(1)}+ L_F^{(2)}$ is\footnote{To show this one has to use the identity that  for  conformal Killing  vector $X$ of weight $\l$,
		$
		\call_X (G_{AD}\G_{BC}^D) = \l  G_{AD}\G_{BC}^D - G_{AD} \pa_B\pa_C X^D.	$}
	\be 
	\d_P L_F = {d \over dt }\left( P  L_F \right).\label{confferm}
	\ee 
	
	The commutator of a special conformal transformation and a supersymmetry (\ref{Nis1Bgauged})
	generates  a new fermionic symmetry:  a conformal supersymmetry. It is  useful to work out how it acts on the fields and to check the closure of the full algebra on all the fields.
	The parameters for the fermionic  symmetries can be conveniently combined into a single  time dependent parameter
	\be
	\S = \e + \h t,
	\ee 
	where  $\e$ is the supersymmetry parameter  and $\h$ parametrizes conformal supersymmetry{; both are time-independent.}
	The fermionic transformations of the fields are
	\begin{eqnarray}
		\delta_\Sigma x^A&=&-i \Sigma  \chi^A, \\
		\delta_\Sigma \chi^A&=&\Sigma D_t x^A+2\dot{\Sigma} \xi_\perp^A\label{sigmachi}\\
		\delta_\Sigma a^I&=&2i \pa_A h^I \dot \Sigma  \chi^A.\label{aferm}
	\end{eqnarray}
	We note in particular  that $a^I$ does transform under  conformal supersymmetry. The full symmetry algebra closes  to a gauged version of the  $ osp (1|2)$ superconformal algebra, with the commutation relations (\ref{gaugedbossum4},\ref{confbossum3},\ref{confbossum4}) supplemented by 
	\bea 
	{}[\delta_{P},\delta_{\Sigma_1}]&=&\delta_{\Sigma_2}\qquad\qquad\mbox{where}\quad \Sigma_{2}=-P\dot \Sigma_{1}+\frac{1}{2}\dot P\Sigma_{1}\\
	{}[\delta_{\Sigma_1},\delta_{\Sigma_2}]&=&\delta_{P}+\delta_\lambda \qquad\ \mbox{where}\quad P=2i\Sigma_{1}\Sigma_{2},\ \ 
	\lambda^I=-2i\left(a^I\Sigma_{1}\Sigma_{2}+h^I\frac{d}{dt}(\Sigma_{1}\Sigma_{2})\right).\nonumber\\
	\eea
	
	Summarized, to make the gauged $\caln = 1B$ sigma model invariant under $osp(1|2)$ superconformal symmetry (SCS) we need   the following structures in addition to (\ref{gaugedbossum1}-\ref{gaugedbossum6},\ref{confbossum1}-\ref{confbossum6}, \ref{gaugedfermsum1}-\ref{gaugedfermsum3}):
	\begin{tcolorbox}[ams align]
		{\rm CS: }& &\d_P\chi^A &= P \dot \chi^A + \dot P \left( \pa_B \x^A \chi^B + \half \chi^A \right) &  \label{conffermsum1} \\
		{\rm SCS: }& & \delta_\Sigma x^A&=-i \Sigma  \chi^A &
		\delta_\Sigma \chi^A&=\Sigma D_t x^A+2\dot{\Sigma} \xi_\perp^A\nonu
		&&	\delta_\Sigma a^I&=2i \pa_A h^I \dot \Sigma  \chi^A &&  \label{conffermsum2}\\
		{\rm A:}&&  {}[\delta_{P},\delta_{\Sigma_1}]&=\delta_{\Sigma_2}&  \Sigma_{2}&=-P\dot \Sigma_{1}+\frac{1}{2}\dot P\Sigma_{1} \label{conffermsum3}\\
		&&	{}[\delta_{\Sigma_1},\delta_{\Sigma_2}]&=\delta_{P}+\delta_\lambda&  P&=2i\Sigma_{1}\Sigma_{2} \label{conffermsum4}\\
		&&&& &\hspace{-1.3cm}	\lambda^I=-2i\left(a^I\Sigma_{1}\Sigma_{2}+h^I\frac{d}{dt}(\Sigma_{1}\Sigma_{2})\right)\nonu
		{\rm GC: }& & \call_\x C_{ABC} &= - C_{ABC} &
		C_{ABC}\x_{\perp}^C &= 2  k_{I [A} \pa_{B]} h^I
		\label{conffermsum5}
	\end{tcolorbox}
	
	\subsection{Noether charges}
	We now compute the Noether charges for the  $ osp (1|2)$ superconformal symmetry. For a symmetry with parameter $\s$ under which the Lagrangian transforms as
	\be
	\d_\s L = {d \over dt} B_{\s}, 
	\ee
	the  expression for the Noether charge, including the fermions, is in our conventions,
	\be 
	\s  Q_{\s} = i^{F_{\s}}\left(  {\pa L \over \pa \dot x^A} \d_{\s} x^A +   {\pa_R L \over \pa \dot \chi^A} \d_{\s} \chi^A- B_{\s}\right) ,\label{chargeformula}
	\ee
	where  the subscript $_R$ means that the fermionic derivative acts from the right in our conventions. The definition includes a factor $i$ for fermionic symmetries to ensure that $ Q_{\s}$ is real.  For later convenience it is useful to write the Noether charges, not in terms of the standard bosonic momentum $p_A$ obtained by varying the Lagrangian while keeping the $\chi^A$ fixed, but in terms of a momentum $\tilde p_A$ arising from varying the Lagrangian while  keeping fixed instead the flat-space-index fermions $\chi^{\underline{A}} = e_A^{\underline{A}} \chi^A$. In other words,
	\be 
	p_A := \left.{\pa L \over \pa \dot x^A}\right|_{\chi^B}, \qquad \tilde p_A := \left. {\pa L \over \pa \dot x^A}\right|_{\chi^{\underline{B}}},
	\ee
	and one finds that the two are related as
	\be
	p_A = \tilde p_A -{i \over 2} \left(  \o_{ABC} - G_{D B} \G^D_{CA}\right)  \chi^B \chi^C.
	\ee
	
	For the gauge generators $M_I$ one finds, using (\ref{Bgauge}),
	\be
	M_I =  k_I^A \left(\tilde p_A - A_A - {i \over 2} \o_{ABC} \chi^B \chi^C  \right) + {i \over 2}  \chi^A \chi^B \nabla_Bk_{IA} - v_I \label{QIferm}
	\ee 
	For the conformal Noether charges we obtain, from (\ref{chargeformula}) and (\ref{confferm}),
	\bea
	H &=& \half \P_A G^{AB} \P_B  + {i \over 2} F_{AB} \chi^A \chi^B+   {1 \over 12} \pa_{[A} C_{BCD]}  \chi^A  \chi^B  \chi^C  \chi^D   + a^I M_I \label{HclNis1}\\
	D&=& \x_\perp^A \P_A+ h^I M_I \label{DclNis1}\\
	K &=& 2 \x_{\perp A} \x_\perp^A,\label{KclNis1}
	\eea
	where  we abbreviated
	\be
	\P_A := \tilde p_A -{i \over 2}  \left( \o_{ABC} - \half C_{ABC}\right)\chi^B \chi^C- A_A.
	\ee
	The supercharge is found to be, using (\ref{Bgauge}),
	\be 
	Q = \chi^A  \left(\tilde p_A - A_A- { i \over 2} \o_{ABC} \chi^B \chi^C  + { i \over 12} C_{ABC} \chi^B \chi^C \right).\label{QclNis1}
	\ee
	This reduces to  the expression in \cite{Michelson:1999zf} for ungauged models.
	The conformal supercharge $S$ could similarly  be found by computing the variation of the Lagrangian. A simpler way is to note from (\ref{conffermsum3}) that it should  equal  the Poisson bracket of the   special conformal and supersymmetry  generators:
	\be 
	S :=  \{  K, Q \}
	=-  2     \chi^A \x_{\perp A}. \label{SclNis1}
	\ee
	In the last line, we have used 
	the canonical Poisson/Dirac brackets
	\bea
	\{ x^A,\tilde  p_B \} &=& \d^A_B\\
	\{ \tilde p_A, \chi^{\underline B} \} &=& 0 \\
	\{ \chi^{\underline A} , \chi^{\underline B} \} &=& - i \d^{\underline A \underline B}.
	\eea
	The last line  arises from a Dirac bracket due to the constraint $\Pi_{\underline A} - i \chi_{\underline A} \approx 0$, where $ \Pi_A := \pa L/ \pa \dot \chi^A$  (the naive Poisson bracket would give twice the right-hand side). 
	The algebra of transformations (\ref{gaugedbossum4},\ref{confbossum3},\ref{confbossum4},\ref{conffermsum3},\ref{conffermsum4}) guarantees that the Poisson brackets of the Noether charges  (\ref{HclNis1}-\ref{HclNis1},\ref{QclNis1},\ref{SclNis1}) weakly form the algebra $osp(1|2)$.  As in the bosonic models, we can just as well work with a redefined dilatation charge where we drop the second term in (\ref{DclNis1}). This is natural from the point of supersymmetry as it is the latter that arises from the Poisson bracket of the supersymmetry and superconformal charges.

	\subsection{Quantization}
	To quantize the theory, we start from operators obeying the canonical commutation relations
	\begin{align}
		[ x^A,\tilde  p_B ] &= i \d^A_B &
		[ \tilde p_A, \chi^{\underline B} ] &= 0 \\
		\{ \chi^{\underline A} , \chi^{\underline B} \} &= \d^{\underline A \underline B}, &\{ \chi^A , \chi^B \} &= G^{AB}\label{ccrferm}
	\end{align}
	A useful consequence of these is
	\be 
	[ \tilde p_A, \chi^B ] = - i (\o_{A \ C}^{\ B \ }- \G^B_{AC} )\chi^C.\label{pchicomm}
	\ee
	In addition   we will assume that 
	these act on  a Hilbert space with an inner product  such that the Hermiticity property (\ref{momquant})  generalizes to
	\bea 
	\tilde p_A^\dagger &=&\tilde p_A - i \G_{AB}^B\label{momquant2}\\
	( \chi^A)^\dagger &=& \chi^A \label{Hermprop}
	\eea
	
	We now proceed to fix the ordering ambiguities in the quantum operators $ M_I, Q, S, H, D, K$ such that they are Hermitian   and (weakly) obey the $osp(1|2)$ commutation relations. We {find} that the appropriate quantum operators are given by
	\begin{tcolorbox}[ams align]
		M_I' &= k_I^A \left( \P_A  - {i \over 4}  C_{ABC} \chi^B \chi^C\right)  - {i \over 2} \chi^A \chi^B  \nabla_A k_{IB}-v_I - {i \over 2} f_{IJ}^{ \ \ J}\label{ospgen1}  \\
		Q &= \chi^A \P_A - {i \over 6} C_{ABC} \chi^A \chi^B \chi^C\label{ospgen2}\\
		S &= - 2 \chi^A \x_{\perp A} \label{ospgen3}\\
		H' &= \half \P_A^\dagger G^{AB} \P_B + {i \over 2} F_{AB} \chi^A \chi^B - \frac{1}{8} C_{ABC} C^{ABC} \nonu & -{1 \over 4}\left( \check R_{ABCD}+ {2 \over 3} \check \nabla_A C_{BCD}+ \half C_{AB}^{\ \ \ E}C_{CDE} \right) \chi^A \chi^B \chi^C \chi^D\label{ospgen4}\\
		D'
		&= \half \xi_{\perp}^A \P_A + {\rm h.c.}\label{ospgen5}\\
		K &= 2 \x_{\perp A} \x_\perp^A\label{ospgen6}\\
		\P_A &:= \tilde p_A - {i \over 2} \left( \o_{ABC} - \half C_{ABC} \right) \chi^B \chi^C - A_A \label{ospgen7}
	\end{tcolorbox}
	We recall  that the prime in  $M_I', H', D'$ means that  these charges  include improvement terms in order to  realize the algebra on the physical Hilbert space. 	The above operators indeed weakly  obey the $osp(1|2)$ commutation relations:
	\begin{align}
		[M_I' , Q ] &= 0, & [M_I', S] &=0\label{osp1}\\
		\{ Q, Q \} &= 2 H' & \{ S, Q \} &=- 2 D'  & & \label{osp2}\\
		\{ S, S \} &= 2 K&  [Q, K ] &= -i S & [Q,D'] &\approx {i\over 2} Q
		\label{osp3}   \\
		[S,D'] &=- {i\over 2}S, & [S,H'] &= iQ &&\label{osp4}\\
		[D',H'] &\approx -i H{'}, &
		[D', K] &= i K &
		[H',K] &= 2 i  D{'}.\label{osp5}
	\end{align} 
	Let us briefly describe how the expressions (\ref{ospgen1}-\ref{ospgen7}) were obtained.
	The operators $ S$ and $K$ are free of ordering ambiguities  and 
	the classical Noether charges lead directly to (\ref{ospgen3},\ref{ospgen6}). The  expression (\ref{ospgen1}) for the gauge generators $M_I'$ follows from the Noether charge (\ref{QIferm}), with the need for the term proportional to $f_{IJ}^{ \ \ J}$ explained in Section \ref{Secbosquant}. Following  \cite{Michelson:1999zf}, we have chosen the operator ordering in the supercharge $Q$ in (\ref{ospgen2}) such that the result is Hermitian. To show the latter property, one needs  (\ref{pchicomm}). The properly ordered Hamiltonian $H'$ and dilatation generator $D'$ in (\ref{ospgen4},\ref{ospgen5}) are then defined by the superalgebra relations (\ref{osp2}). We also made use of the following identities
	\bea
	\, [\P_A, \chi^B] &=& i \check \G^B_{\ AC} \chi^C \label{qids1} \\
	\, [\P_A, \P_B] &=& - \half \check R_{ABCD} \chi^C \chi^D + i F_{AB} \label{qids2} \\
	\,[\P_A , t_{B_1 \ldots B_n}  \chi^{B_1}\ldots  \chi^{B_n}] &=&- i \check \nabla_A  t_{B_1 \ldots B_n}  \chi^{B_1} \ldots  \chi^{B_n} \label{qids3} \\
	\,[M_I , t_{A_1 \ldots A_n}  \chi^{A_1}\ldots  \chi^{A_n}] &=&- i \call_{k_I} t_{A_1 \ldots A_n}  \chi^{A_1} \ldots  \chi^{A_n} \label{qids4}
	\eea
	To verify the algebra $osp(1|2)$ commutation relations, it suffices to check the relations (\ref{osp3}), the remaining ones then follow from (super-) Jacobi identities.    	
	
\section{ Gauged sigma models with $ su (1,1|1)$ superconformal symmetry}\label{SecNis2}
In this section we comment on models with $\caln=2B$ super(conformal) symmetry. For the sake of brevity we will focus on a specific class of models that contain an R-symmetry and which furthermore can be extended to the $\caln=4B$ models of the next section, but we do point out possibilities for generalizations of which we leave a full classification to further investigation.

Given some extra conditions specified below, the supersymmetry of the ungauged Lagrangian (\ref{L1ferm}, \ref{L2ferm}) is extended to the $\caln=2B$ susy transformations
	\bea
	\d_{\e,\tilde{\e}} x^A &=& - i \left(\e \chi^A + \tilde{\e} J^A{}_B \chi^B \right) \\
	\d_{\e,\tilde{\e}} \chi^A &=&  \left( \epsilon \dot x^A  - \tilde{\e} \left( J^A{}_B \dot{x}^B + i \partial_C J^A{}_B \chi^C \chi^B \right) \right)\label{Nis2Bgl}
	\eea where $\tilde{\e} = \tilde{\e}^*$ is another real fermionic parameter. Although this is not strictly necessary, we assume the presence of a $U(1)$ ${R}$-symmetry as well. Models with $\caln=2$ susy but without such $R$ symmetry exist and satisfy slightly weaker conditions than the ones we'll derive, see e.g. \cite{Gibbons:1997iy}. Since upon the addition of conformal symmetry the presence of $R$-symmetry becomes mandatory we find it convenient to include it from the beginning. The $R$-symmetry transformation includes an a priori unconstrained vector field $\omega^A$ (which could in some models be zero) and is more explicitly given by
	\bea
	\delta_r x^A &=& -r \o^A  \label{ugt1g} \\ \delta_r \chi^A &=& - r \left( \partial_B \o^A + \frac{1}{2} J^A{}_B \right) \chi^B. \label{ugt3g}
	\eea  Then the closure of the algebra 
	\bea
	~[\delta_r,\delta_r] &=& 0 \\
	~[\delta_r,\delta_{\e_1,\tilde{\e}_1}] &=& \delta_{\e_2,\tilde{\e}_2} \qquad\qquad\mbox{where}\quad \e_{2}=-\frac{1}{2} \tilde{\e}_1\quad,\quad \tilde{\e}_2 = \frac{1}{2} \e_1 \\  ~[\delta_{\e_1,\tilde{\e}_1},\delta_{\e_2,\tilde{\e}_2}] &=& u \pa_t \qquad\qquad~~\mbox{where}\quad u= 2i \left( \e_1 \e_2 + \tilde{\e}_1 \tilde{\e}_2 \right)
	\eea requires  \be  J^A{}_C J^C{}_B = - \delta^A_B,  \quad\quad J^D{}_{[B} \partial_{|D|} J^A{}_{C]} - J^A{}_D \partial_{[B} J^D{}_{C]} = 0, \quad\quad \call_\o J^A{}_B = 0, \label{jj} \ee where the second condition is the Nijenhuis integrability condition, and hence $J^A{}_B$ is an integrable complex structure on the target manifold, with respect to which $\omega$ needs to be holomorphic.
		Note that invariance under $\caln=1$ supersymmetry together with invariance under $R$-symmetry guarantees invariance under $\caln=2$ supersymmetry via the algebra.
		
		Furthermore, invariance\footnote{The Lagrangian transforms with a total derivative term as given below in (\ref{drL}).} of the ungauged model (\ref{L1ferm},\ref{L2ferm}) under the $R$-symmetry is equivalent to the geometric conditions
	\begin{align}
	F_{AC}(J)^C{}_B+F_{CB}(J)^C{}_A&=0, &	G_{AC}(J)^C{}_B+G_{CB}(J)^C{}_A&=0, \label{FGhermit} \\ \call_{\o} G_{AB} &= 0, & i_\o F &= 0,  \label{ung4}\\ \check{\nabla}_A J_{BC}-\call_{\o} C_{ABC} &= 0, &  J^B{}_{[E} \partial_A C_{BCD]} &= 0. \label{JpacNis2}
	\end{align}
 We recognize the first line as the requirement of Hermiticity of both the field strength and metric, while the second line demands both the metric and external gauge field strength to be invariant along the vector field $\omega$. Finally the most interesting condition is the first equation in \eqref{JpacNis2}. First note that by anti-symmetry it implies $\hat\nabla_{(A}J_{B)C}=0$, which is the slightly weaker condition one obtains by requiring only invariance under $\caln=2$ supersymmetry (without $R$-symmetry), as found in e.g. \cite{Gibbons:1997iy, Michelson:1999zf}. In what follows we will restrict attention to the special case
	\begin{equation}
	   \call_\omega C_{ABC}=0\qquad \check \nabla_A J_{BC}=0\,,\label{N2spec}
	\end{equation} 
	The equality on the right expresses the condition that $\check \nabla$, see \eqref{torsionder}, must be the Bismut\footnote{The Bismut connection is the unique connection with totally anti-symmetric torsion tensor preserving a given Hermitian structure, see e.g \cite{Fedoruk:2014jba} for a pedagogic discussion.} connection for the Hermitian structure\footnote{More generically, if $\call_\omega C_{ABC}\neq 0$ the first constraint in \eqref{JpacNis2} determines the connection $\check \nabla=\nabla_{\mathrm{Bismut}}+\calb$ to be some deformation of the Bismut connection, that needs to satisfy $\call_\o \calb_{ABC}=\calb^D{}_{A[B}J_{C]D}$.} ($J^A{}_B, G_{AB}$).  The explicit expression for $C_{ABC}$ is then given by \cite{Fedoruk:2014jba}
	\begin{equation}
	C_{ABC} = - 3 J^D{}_A J^E{}_B J^F{}_C \nabla_{[D} J_{EF]} \label{CJbism} 
	\end{equation} 
	Note that in the special case \eqref{N2spec} the second condition in \eqref{JpacNis2} is automatically satisfied.
	
	As in the $\caln=1$ case we can proceed and gauge some isometries. The gauge-covariantized supersymmetry transformations are
	\bea
	\d_{\e,\tilde{\e}} x^A &=& - i \left(\e \chi^A + \tilde{\e} J^A{}_B \chi^B \right) \\
	\d_{\e,\tilde{\e}} \chi^A &=&  \left( \epsilon \dot D_tx^A  - \tilde{\e} \left( J^A{}_B D_t{x}^B + i \partial_C J^A{}_B \chi^C \chi^B \right) \right) \label{Nis2Bgl2}
	\eea and the gauge-covariantized $R$-symmetry is given by
	\bea
	\delta_r x^A &=& -r \o^A  \label{t11g} \\ \delta_r a^I &=& r V_A^I D_t x^A\label{t21g} \\ \delta_r \chi^A &=& - r \left( \partial_B \o^A + V_B^I k_I^A + \frac{1}{2} J^A{}_B \right) \chi^B \label{t31g}
	\eea
	where the deformation upon gauging is captured by the one-forms
	\begin{equation}
		V^I_A = J^B{}_A \partial_B h^I .
	\end{equation} 
	
	The algebra of transformations closes into
	\bea 
	~[\delta_{\l},\delta_{r}]&=&0 \nonu
	~[\delta_r,\delta_{\e_1,\tilde{\e}_1}] &=& \delta_{\e_2,\tilde{\e}_2}+\delta_\lambda \quad\mbox{where}\quad \e_{2}=\frac{1}{2} \tilde{\e}_1~,~ \tilde{\e}_2 = - \frac{1}{2} \e_1~ \nonu  ~[\delta_{\e_1,\tilde{\e}_1},\delta_{\e_2,\tilde{\e}_2}] &=& u \pa_t + \delta_\lambda \quad~~\mbox{where}\quad u= 2i \left( \e_1 \e_2 + \tilde{\e}_1 \tilde{\e}_2 \right)~,\quad~\lambda^I= -2i a^I \left(\e_1 \e_2 + \tilde{\e}_1 \tilde{\e}_2 \right) ,\nonumber 
	\eea provided the following  constraints hold
	\begin{align}
		\partial_{[A} V^I_{B]} &= 0, & \call_{k_I} J^A{}_B &= 0, \qquad   [\o,k_I]^A = 0, \label{Scond01} \\ 
		\call_\o \left(J\right)^A{}_B &= V^I_C J^C{}_B k_I^A - V^I_B J^A{}_C k_I^C, &  \call_\o \left(dh^I\right)_A &= \left(\partial_A h^J V^I_B - V_A^J \partial_B h^I\right)k_J^B .\label{2liodh}
	\end{align}
	
	The $\caln=1B$ action (\ref{gaugedbossum1},\ref{gaugedfermsum1}) remains invariant under these gauge-covariantized $\caln=2B$ susy transformations. We check this by deriving that the Lagrangian transforms under the $U(1)$ $R$-symmetry (\ref{t11g},\ref{t31g}) as 
	\be
	\delta_r L = -r \frac{d}{dt} \left(i_\o A \right) \label{drL}
	\ee
if one imposes the additional constraints\footnote{In case $\calb_{ABC}\neq 0$ the conditions (\ref{bismutcond}, \ref{Scondxx1}) get replaced by the weaker condition $\check{\nabla}_A J_{BC} = \call_{\o} C_{ABC} +3V^I_{[A} C_{BC]D} k_I^D$ .} : 
	\begin{eqnarray}
		\check{\nabla}_A J_{BC} &=& 0\label{bismutcond} \\
		\call_{\o} C_{ABC}&=&- 3V^I_{[A} C_{BC]D} k_I^D \label{Scondxx1} \\ \call_{\o} G_{AB} &=& - 2 k_{I(A}V^I_{B)} \label{Scond00} \\ i_\o F &=& v_I V^I_A \label{Sl1g1} \\  \call_{\o} v_J &=& - k_J^A V_A^I v_I \label{Sl1g3}
	\end{eqnarray}

	We see that geometric constraints (\ref{Scondxx1}-\ref{Sl1g1}) have been slightly deformed compared to their ungauged counterparts (\ref{ung4}-\ref{JpacNis2}).

	To summarize, ${R}$-symmetry (denoted as (RS) below) leads to the following structures in addition to (\ref{gaugedbossum1}-\ref{gaugedbossum6}), and (\ref{gaugedfermsum1}-\ref{gaugedfermsum3})
	\begin{tcolorbox}[ams align]
		{\rm {R}S: }& &\delta_{{r}} x^A &= -r \o^A \qquad,\qquad \d_{{r}} a^I =  r V^I_A D_t x^A \nonu & & \d_{{r}} \chi^A &= - r \left(\partial_B \o^A + V_B^I k_I^A + \frac{1}{2} J^A{}_B \right) \chi^B \label{gaugedNis2sum0} \\ {\rm A:} & &    
		~[\delta_{\l},\delta_{r}]&=0 \nonu
		&& ~[\delta_r,\delta_{\e_1,\tilde{\e}_1}] &= \delta_{\e_2,\tilde{\e}_2}+\delta_\lambda ~,\quad \e_{2}=\frac{1}{2} \tilde{\e}_1~,~ \tilde{\e}_2 = - \frac{1}{2} \e_1~ \nonu && ~[\delta_{\e_1,\tilde{\e}_1},\delta_{\e_2,\tilde{\e}_2}] &= u \pa_t + \delta_\lambda \quad,\quad u= 2i \left( \e_1 \e_2 + \tilde{\e}_1 \tilde{\e}_2 \right)~,~\lambda^I= -2i a^I \left(\e_1 \e_2 + \tilde{\e}_1 \tilde{\e}_2 \right)  \\ 
		{\rm SC:}& & J^A{}_C J^C{}_B& =-\delta^A_B \quad,\quad J^D{}_{[B} \partial_{|D|} J^A{}_{C]} - J^A{}_D \partial_{[B} J^D{}_{C]} = 0\label{gaugedNis2sum1}  \\
		&&  \call_{k_I} J &= 0  \qquad\qquad [\o,k_I]^A = 0 \qquad\qquad \partial_{[A} V^I_{B]} = 0 \label{gaugedNis2sum2} \\  && \call_\o \left(J\right)^A{}_B &= V^I_C J^C{}_B k_I^A - V^I_B J^A{}_C k_I^C \label{gaugedNis2sum7} \\ && \call_\o \left(dh^I\right)_A &= \left(\partial_A h^J V^I_B - V_A^J \partial_B h^I\right)k_J^B \label{gaugedNis2sum8} \\\\
		{\rm GC:}&& 0&= F_{AC}(J)^C{}_B+F_{CB}(J)^C{}_A \label{gaugedNis2sum3}\\
		&& 0&= G_{AC}(J)^C{}_B+G_{CB}(J)^C{}_A \label{gaugedNis2sum33}\\
		&& \check \nabla_A J_{BC}&=0\\
		&&  \call_\o C_{ABC} &= -3 V^I_{[A} C_{BC]D} k_I^D  \label{gaugedNis2sum4} \\ && \call_\o G_{AB} &= - 2k_{I(A} V^I_{B)} \qquad\qquad i_\o F = v_I V_A^I \label{gaugedNis2sum5} \\&& \call_\o v_J &= - k_J^A V_A^I v_I \label{gaugedNis2sum6} 
	\end{tcolorbox}

	\subsection{Conformal invariance}
	
	The superconformal transformations of the fields are
	\begin{eqnarray}
		\delta_\Sigma x^A&=&-i(J^\alpha)^A{}_B\Sigma_\alpha \chi^B\qquad\qquad\qquad\qquad  J^\alpha=(J,\mathds{1}), \bar J^\alpha=(-J,\mathds{1})\qquad\\
		\delta_\Sigma a^I&=&2i V_A^{\alpha\, I}\dot \Sigma_\alpha \chi^A\label{afermg}\quad~~\quad\qquad\qquad\qquad V_A^{\alpha\, I}=\partial_Bh^I (J^\alpha)^B{}_A\\
		\delta_\Sigma \chi^A&=&(\bar J^\alpha)^A{}_B\left(\Sigma_\alpha D_t x^B+2\dot{\Sigma}_\alpha \xi_\perp^B\right)+i\partial_C (J^\alpha)^A{}_B\Sigma_\alpha \chi^C\chi^B\label{ssigmachi} 
	\end{eqnarray} 
	Here, for notational simplicity and future convenience, 
	 we have defined a two-component index $\a$ taking the values  $\alpha=(3,4)$. 
The supersymmetry parameters are relabeled as $\epsilon^\alpha = (\epsilon^3,\epsilon^4) \equiv (\tilde{\epsilon},\epsilon)$ and  the superconformal parameters are defined as  $\Sigma^\a=\epsilon^\a+t\, \eta^\a$.

	The full symmetry algebra closes  to a gauged version of the  $ su (1,1|1)$ superconformal algebra, with the commutation relations (\ref{confbossum3},\ref{confbossum4},\ref{conffermsum3},\ref{conffermsum4}) being supplemented by 
	the additional commutators listed in (\ref{rnis2sum3},\ref{rnis2sum4},\ref{rnis2sum5},\ref{rnis2sum6}) below. In particular, the commutator (\ref{rnis2sum4}) fixes $\omega^A$ as 
	\be
	\omega^A=-(J)^{A}{}_B\xi^B_\perp \label{omegaconf} \ee

	We note that with this choice of $\omega^A$ in (\ref{rnis2sum1}), the constraints (\ref{Scond01}-\ref{2liodh}), (\ref{Scond00}-\ref{Sl1g3}) involving $\o^A$ become a direct consequence of the previously obtained gauge and conformal invariance conditions. In other words, once the conformal symmetry is introduced, the invariance under $R$-symmetry is readily implied both in the ungauged and gauged cases.

	Summarized, making the gauged $\caln = 2B$ sigma models invariant under $su(1, 1|1)$ superconformal symmetry (SCS) leads to the following structures in addition to (\ref{gaugedbossum1}-\ref{gaugedbossum6}), (\ref{gaugedfermsum1}-\ref{gaugedfermsum3}), (\ref{conffermsum1}-\ref{conffermsum5}) and (\ref{gaugedNis2sum0}-\ref{gaugedNis2sum6})
	\begin{tcolorbox}[ams align]
		{\rm SCS: }& & \delta_\Sigma x^A&=-i(J^\alpha)^A{}_B\Sigma_\alpha \chi^B ~
		\qquad\qquad\qquad	\delta_\Sigma a^I=2i V_A^{\alpha\, I}\dot \Sigma_\alpha \chi^A  \nonu && \delta_\Sigma \chi^A&=(\bar J^\alpha)^A{}_B\left(\Sigma_\alpha D_t x^B+2\dot{\Sigma}_\alpha \xi_\perp^B\right)+i\partial_C (J^\alpha)^A{}_B\Sigma_\alpha \chi^C\chi^B \label{rnis2sum2}\\ 
		{\rm A:}&&  {}[\delta_{P},\delta_{\Sigma_1}]&=\delta_{\Sigma_2} \qquad\qquad\qquad \Sigma_{2}=-P\dot \Sigma_{1}+\frac{1}{2}\dot P\Sigma_{1} \label{rnis2sum3} \\
		&&	{}[\delta_{\Sigma_1},\delta_{\Sigma_2}]&=\delta_{P}+ \d_r +\delta_\lambda \qquad ~P=2i\Sigma_{1}\Sigma_{2}\nonu
		&&& \hspace{3.4cm}	\lambda^I=-2i\left(a^I\Sigma_{1}\Sigma_{2}+h^I\frac{d}{dt}(\Sigma_{1}\Sigma_{2})\right)\nonu &&& \hspace{3.3cm}~~r= 2i \epsilon_{\rho \sigma} \Sigma_{1\rho} \dot{\Sigma}_{2\sigma} \label{rnis2sum4} \\	&& {}[\delta_{{r}},\delta_{\Sigma_1}] &= \delta_{\Sigma_2} + \delta_\lambda \qquad\quad~ \Sigma_{2\rho} = \frac{1}{2} r \left( \delta_{\rho 3} \Sigma_{14} - \delta_{\rho 4} \Sigma_{13} \right) \label{rnis2sum5} \\ 	&& {}[\delta_{{r}},\delta_{\lambda}] &= 0 
		\label{rnis2sum6} \\{\rm SC: } & & \o^A &= -J^A{}_B \xi_\perp^B  \qquad,\qquad \call_\xi J^A{}_B = 0 \label{rnis2sum1} 
	\end{tcolorbox} 
	
	\subsection{Noether charges}
	
	Extending the superconformal symmetry from $osp(1|2)$ to $ su (1,1|1)$ brings in three more Noether charges. Of special interest is the Noether charge for the $U(1)$ $R$-symmetry is obtained from (\ref{drL}) by using the general formula (\ref{chargeformula}) as 
	\bea
R &=& - \o^A \left( \Pi_A - \frac{i}{4} C_{ABC} \chi^B \chi^C \right) + \frac{i}{2}  \left( \nabla_B \o_C \right) \chi^B \chi^C - \frac{i}{4} \left( 2 k_{IB} V^I_C + J_{BC} \right) \chi^B \chi^C \nonumber \\ &=& -\left( \o^A \Pi_A - i  \nabla_A \o_B  \chi^A \chi^B \right) \label{Rn2l2m}
	\eea
Also, to write (\ref{Rn2l2m}) we used the identity
	\be 
	\o^{A}C_{ABC} = J_{BC} + 2 \nabla_{[B} \o_{C]} + 2 k_{I[B} V^{ I}_{C]}
	\ee which is implied by (\ref{CJbism}) and (\ref{omegaconf}).
	The Noether charges for the additional supersymmetry and superconformal symmetry can be obtained by computing Poisson brackets with the charges already found. We will directly compute their quantum expressions presently.

	\subsection{Quantization}
	We now address the issue of the operator ordering in the quantum generators of $su(1,1|1)$.  The canonical (anti-)commutators and Hermiticity properties of the fields were given in (\ref{ccrferm}) and (\ref{Hermprop}). We  {find} that the quantum $su(1,1|1)$ generators are given by
	\begin{tcolorbox}[ams align]
		Q^3 &= \chi^A J_{\ \ A}^{B} \P_B + {i \over 2}  J_{[A}^{\ \ D} C_{BC]D} \chi^A \chi^B \chi^C\label{sugen1} \\
		Q^4 &= \chi^A \P_A - {i \over 6} C_{ABC} \chi^A \chi^B \chi^C\label{sugen2}\\
		S^3 &= - 2 \o_A  \chi^A  \label{sugen3},& S^4 &=  - 2 \x_{\perp A}\chi^A  \\
	R &= - \left( \o^A \Pi_A - \frac{i}{2} \nabla_A \o^A - i  \nabla_{[A} \o_{B]} \chi^A \chi^B \right)\label{sugen4}
	\end{tcolorbox}
	where the expressions for $\P_A$  and the remaining generators $M_I', H', D', K$ are the same as in the $osp(1,2)$ case, see (\ref{ospgen1},\ref{ospgen4}-\ref{ospgen7}). We note from (\ref{sugen2},\ref{sugen3}) that $Q^4$ and $S^4$ generate an $osp(1|2)$ subalgebra and therefore satisfy the commutation relations  (\ref{osp1}-\ref{osp3})  already computed in Section \ref{SecNis1}.
One furthermore  verifies the anticommutation relations 
	\begin{align}
	\{ Q^\a , Q^\b \} &= 2 \d^{\a\b} H' , &	\{ S^\a , S^\b \} &= 2 \d^{\a\b} K \\
		 \{ Q^\a, S^\b \}  &=  - 2 \d^{\a\b} D'- 2 \e^{\a\b} R,
	 	\end{align} 
	 	where $\e^{34}\equiv 1$.
The remaining	(gauged) $su(1,1|2)$ commutation relations follow from the  these
elementary ones
upon applying the	 (super-)Jacobi identities.

	The generators simplify when expressed in complex coordinates $(z^m, \bar z^{\bar m})$ adapted to  the complex structure $J$, i.e
	\be 
	J_m{}^{n} = - i \d_m ^n, \qquad J_{\bar m}{}^{\bar n} =  i \d_{\bar m}^{\bar n}\label{complcoordNis2}
	\ee 
	We define the complex super(conformal) charges
	\begin{align} 
		 \calq_- &= \half (-Q^3 + i Q^4), &
	\cals_- &= \half (-S^3 + i S^4).
	\end{align}
These can be written as
	\begin{align}
	\calq_- &=  \chi^{\bar m} \left( i \tilde p_{\bar m} + \o_{\bar m \bar p  n} \chi^{\bar p} \chi^n- i  A_{\bar m}+ {1 \over 8} \pa_{\bar m} \ln \det G \right), &
	\cals_- &= i \chi^{\bar m} \x_{\perp \bar m}. \label{Nis2gencc}
	\end{align}
	To derive these we have made use of some some properties of the Bismut connection expressed in complex coordinates (\ref{complcoordNis2}),
	\bea
	\o_{mnp} &=&   \o_{\bar m \bar n \bar p} =  C_{mnp} = C_{\bar m \bar n \bar p}=0 \\
	\o_{m \bar n \bar p} &=&  \half C_{m \bar n \bar p} , \qquad
	\o_{\bar m  n  p} =  \half C_{\bar m  n  p} \\
	C_{\bar m n \bar p} G^{n \bar p} &=&  2 \G^{ n}_{ n \bar m}  , \qquad
	\o_{\bar m n \bar p} G^{n \bar p} = {1 \over 4}\pa_{\bar m} \ln \det G - \G^n_{n \bar m} .
	\eea
Also important for our purposes is the expression for the gauge generators in complex coordinates,
\bea 
M_I'&=& k_I^A (\tilde p_A - i \o_{A\bar m n} \chi^{\bar m}\chi^n) - i \nabla_{\bar m}k_{In} \chi^{\bar m}\chi^n\nonu
&& - i_{k_I} A- v_I -{i \over 2} f_{IJ}^{\ \ J}+ {i \over 2} \left(\pa_{[\bar n} k_{p]} + k^A \o_{A \bar n  p}\right)G^{\bar n p}\label{MIcc}
\eea
In deriving this expression, use has been made of the Hermiticity property
\be
J^C_{\ A} N_{ICB}+ J^C_{\ B} N_{IAC} =0, \qquad N_{IAB} \equiv \check \nabla_A k_{IB} + C_{ABC}k_I^C.
\ee
	\subsection{Representation on differential forms}\label{Secdiffform}
	As is familiar in supersymmetric sigma models, the symmetry generators can  be linked more directly to the differential geometry of the target space manifold. 
	For this it is useful to represent  the algebra of canonical (anti-) commutation relations (\ref{ccrferm}) on the Hilbert space of antiholomophic differential forms. 
	
	We start by introducing, for the complex coordinates   introduced in (\ref{complcoordNis2}),  a unitary frame satisfying $ e_m^{\underline {\bar m}} =  e_{\bar m}^{\underline m} =0$ and
	\be
	G_{m \bar n} = e_m^{\underline m}  e_{\bar n}^{\underline {\bar n}} \d_{\underline m\underline {\bar n}}
	\ee
	Any $(0,q)$ form $\l_{(q)}$ can be expanded in the unitary frame as follows:
	\be
	\l_{(q)} = {1\over q!} \l_{ \underline {\bar m_1} \ldots  \underline {\bar m_q}} e^{ \underline {\bar m_1}}\wedge \ldots \wedge e^{ \underline {\bar m_q}}
	\ee
	The inner product on the space of antiholomorphic forms is taken to be
	\be 
	(\k, \l) = \sum_q \int \bar \k_{(q)}^{ \underline m_1 \ldots \underline m_q} \l_{ (q) \underline {\bar m_1} \ldots  \underline {\bar m_q}} \sqrt{G} |d^n z|^2 \label{inprod}
	\ee
	where $g \equiv \det \{ G_{m \bar n} \}$.
	We define the momentum and fermion operators to act on the Hilbert space as follows:
	\bea
	\tilde p_m \l_{(q)}  &=& - i \pa_m \l_{(q)}:=  -{ i\over q!}\pa_m \l_{ \underline {\bar m_1} \ldots  \underline {\bar m_q}} e^{ \underline {\bar m_1}}\wedge \ldots \wedge e^{ \underline {\bar m_q}}\\
	\tilde p_{\bar m} \l_{(q)}  &=& - i \pa_{\bar m} \l_{(q)}:= -{ i\over q!}\pa_{\bar m} \l_{ \underline {\bar m_1} \ldots  \underline {\bar m_q}} e^{ \underline {\bar m_1}}\wedge \ldots \wedge e^{ \underline {\bar m_q}}\label{pbar}\\
	\chi^{\underline m}  &=& {\delta \over \delta e^{\underline {\bar m}}}\\
	\chi^{\underline {\bar m}} &=&  e^{\underline {\bar m}} \wedge 
	\eea 
	One checks that these realize the canonical (anti-)commutation relations  (\ref{ccrferm}) in the complex basis:
	\be [ z^m ,\tilde  p_n ] =  i \d^{ m}_{ n}, \qquad 
	[ \bar z^{\bar m} ,\tilde  p_{\bar n} ] =   i \d^{\bar m}_{\bar n},\qquad  \{ \chi^{\underline m} , \chi^{\underline {\bar n}} \} = \d^{\underline m \underline {\bar n}}.
	\ee
	and that (\ref{inprod}) leads to the correct Hermiticity properties (\ref{Hermprop})
	\be
	(\tilde p_m)^\dagger = \tilde p_{\bar m} - i \G_{\bar m \bar n}^{\bar n} , \qquad
	( \chi^{\underline m})^\dagger = ( \chi^{\underline {\bar m}}).
	\ee
	A useful property of this realization is
	\be 
	( i \tilde p_{\bar m} + \o_{\bar m  \bar p n} \chi^{\bar p}  \chi^n )\l = \nabla_{\bar m}\l.
	\ee

	Using  (\ref{Nis2gencc},\ref{MIcc}) one then finds that   the  (super-)conformal charges $\calq_-, \cals_-$  and the gauge generators $M_I'$ are  realized as
	\begin{tcolorbox}[ams align]
		\calq_- &= \bar \pa +\left(  {1 \over 8}\bar \pa \ln \det G  - i A_{(0,1)}  \right) \wedge \label{Nis2diff1}\\
		\cals_- &= -{i \over 2} (\bar \pa K)\wedge\label{Nis2diff2}\\
			M_I'&= -i \call_{k_I} - (i_{k_I} A + v_I) -{i \over 2} f_{IJ}^{\ \ J}+ {i \over 2} (\pa_{[\bar n} k_{p]} + k^A \o_{A \bar n  p})G^{\bar n p}\label{Nis2diff0}
	\end{tcolorbox}
	We see from (\ref{Nis2diff1}) that the supercharge $\calq_-$ is realized as a twisted Dolbeault operator. Its nilpotency  follows from the vanishing of $F_{(0,2)}$ which  in turn follows from (\ref{FGhermit}).
	This realization (\ref{Nis2diff0}-\ref{Nis2diff2}) allows for a differential geometric  description of  the quantum mechanical Hilbert space. For example, supersymmetric ground states belong to cohomology  classes of the twisted Dolbeault operator (\ref{Nis2diff1}).
	
	In addition, physical states should be annihilated by the gauge generators $M_I'$ in (\ref{Nis2diff0}). The Lie derivative term acts within the Hilbert space thanks to the holomorphicity of the $k_I$. As before, the remaining terms in this expression show that the wavefunctions should be seen as  sections of an appropriate bundle. The last term signifies an additional transformation by a  phase   which was absent in the bosonic models; the geometric origin of this phase is a small but interesting question which we leave open at present.

	\section{ Gauged sigma models with $ D(2,1;\a)$ superconformal symmetry}\label{SecNis4}
	In this section we construct gauged sigma models with $\caln = 4B$ supersymmetry and derive the conditions for this to be extended to $ D(2,1;\a)$ superconformal symmetry. As in the previous cases we will also cosntruct the quantum symmetry generators and their geometric realization on the Hilbert space of anitholomorpic forms. 
	
	\subsection{The gauged $\caln = 4B$ supersymmetric sigma model}
	Let us first review the conditions imposed by $\caln = 4B$ supersymmetry on ungauged sigma models with Lagrangian (\ref{Lbos},\ref{L1ferm},\ref{L2ferm}), referring to \cite{Gibbons:1997iy} for more details.
	The  $\caln = 4B$ Poincar\'{e} supersymmetry transformations are parametrized by 4 real fermionic parameters $\e^\r, \r= 1, \ldots , 4$ and act as
	\begin{eqnarray}
		\delta_\epsilon x^A&=&-i(J^\rho)^A{}_B\epsilon_\rho \chi^B\\
		\delta_\epsilon \chi^A&=&(\bar J^\rho)^A{}_B\e_\rho \dot x^B+i\partial_C (J^\r)^A{}_B\epsilon_\r \chi^C\chi^B\label{eopsilonchiung}
	\end{eqnarray}
We will consider theories which in  addition possess  an $su(2)$ $\tilde R$-symmetry with parameters $\tilde r^i, i = 1,2,3$ acting as
	\begin{eqnarray}
		\delta_{\tilde r}x^A&=&0\\
		\delta_{\tilde r}\chi^A&=&\frac{1}{2}\tilde r^i (J^i)^A{}_B\chi^B.\label{rtildeung}
	\end{eqnarray}
	In the above expressions  we have defined
	\be
	J^\r = (J^i,\mathds{1}), \qquad 
	\bar J^\r=(-J^i,\mathds{1}), \qquad i = 1,2,3,\label{Jrhodef}
	\ee
	and the closure of the algebra requires $J^i, i = 1,2,3$  to form an integrable quaternionic structure:
	\begin{equation}
		(J^i)^A{}_C (J^j)^C{}_B=-\delta^{ij}\delta^A_B+\epsilon_{ijk}(J^k)^A{}_B,\qquad \caln(J^i,J^j)^A_{BC}=0, \label{qstruct}
	\end{equation}
	where we introduced the Nijenhuis concomitant
	\begin{equation}
		\caln(J^i,J^j)^A_{BC}\equiv(J^{(i})^D{}_{[B}\partial_{|D|} (J^{j)})^A{}_{C]}-(J^{(i})^A{}_{D}\partial_{[B} (J^{j)})^D{}_{C]}. \label{Ndef}
	\end{equation} 
	
	The geometric conditions\footnote{Requiring invariance only under Poincar\'{e} supersymmetry (\ref{eopsilonchiung}), without imposing the $\tilde R$-invariance (\ref{rtildeung}) leads to a slightly weaker set of conditions spelled out in \cite{Gibbons:1997iy}.} on the target space in order for the action (\ref{Lbos},\ref{L1ferm},\ref{L2ferm}) to be invariant reduce to
	\begin{eqnarray}
		F_{AC}(J^i)^C{}_B+F_{CB}(J^i)^C{}_A&=&0\label{Nis4ung1}\\
		G_{AC}(J^i)^C{}_B+G_{CB}(J^i)^C{}_A&=&0\label{Nis4ung2}\\
		\check \nabla_A (J^i)^B{}_{C}&=&0\label{Nis4ung3}.
	\end{eqnarray}
	The first two relations state that the field strength $F_{AB}$ and metric $G_{AB}$ are simultaneously Hermitian with respect to all three complex structures $J^i$.  We remark that (\ref{Nis4ung3}) implies that the four-form $dC$ is also Hermitian, \be \partial_{[E}C_{BCD]}(J^i)^E{}_A=0,\ee  which will be needed below when proving invariance of the action. The last condition (\ref{Nis4ung3}) means that the three different complex structures are covariantly constant with respect to the {\em same} torsionful covariant derivative (\ref{checknabladef}). The last two conditions  (\ref{Nis4ung2},\ref{Nis4ung3}) define a {\em weakly hyperK\"{a}hler with torsion (wHKT) manifold} \cite{Gibbons:1997iy}. 
	
	Now let us consider  gauged $\caln=4B$ sigma models  where, as before,  the fields transform under gauge transformations of the form   (\ref{gaugedbossum2}) determined by a set of vector fields $k_I$  generating  a $\mathrm{G}$-action as in  (\ref{gaugedbossum3}). The closure of the combined algebra of gauge and supersymmetry transformations requires the $k_I$ to be tri-holomorphic:
	\be 
	\call_{k_I} J^i = 0, \qquad i = 1,2,3.\label{ktrihol}
	\ee
	One then shows that, if the conditions\footnote{One can show that the condition  $\call_{k_I} C =0 $ (see (\ref{gaugedfermsum3})) on the fermionic Lagrangian follows from (\ref{gaugedbossum5},\ref{ktrihol}) and the supersymmetry condition (\ref{Nis4ung3}).}  (\ref{gaugedbossum5},\ref{gaugedbossum6}) for gauging the bosonic  Lagrangian are satisfied, the model is invariant under the gauge-covariantized supersymmetry transformations
	\begin{eqnarray}
		\delta_\epsilon x^A&=&-i(J^\rho)^A{}_B\epsilon_\rho \chi^B\nonu
		\delta_\epsilon a^I&=&0\nonu
		\delta_\epsilon \chi^A&=&(\bar J^\rho)^A{}_B\e_\rho D_t x^B+i\partial_C (J^\r)^A{}_B\epsilon_\r \chi^C\chi^B\label{eopsilonchi}
	\end{eqnarray}
	as well as under $\tilde R$-symmetries:
	\begin{eqnarray}
		\delta_{\tilde r}x^A&=&0\nonu
		\delta_{\tilde r}a^I&=&0\nonu
		\delta_{\tilde r}\chi^A&=& \frac{1}{2}\tilde r^i (J^i)^A{}_B\chi^B.\label{rtildegauged}
	\end{eqnarray}
	Note that the gauge fields $a^I$ are  once again inert under these transformations.
	
	Summarized,  $\caln=4B$ supersymmetry  and  gauge invariance require the following structural (SC) and geometric conditions (GC)  on the target space in addition to (\ref{gaugedbossum1}-\ref{gaugedbossum6}) and (\ref{gaugedfermsum1}-\ref{gaugedfermsum3})
	\begin{tcolorbox}[ams align]
		{\rm SC:}&& (J^i)^A{}_C (J^j)^C{}_B& =-\delta^{ij}\delta^A_B+\epsilon_{ijk}(J^k)^A{}_B, & \caln(J^i,J^j)^A_{BC} &=0\label{gaugedNis4sum1}  \\
		&&  \call_{k_I} J^i &= 0 &&&\label{gaugedNis4sum2}\\
		{\rm GC:}&& 0&= F_{AC}(J^i)^C{}_B+F_{CB}(J^i)^C{}_A, &&&\label{gaugedNis4sum3}\\
		&& 0&= G_{AC}(J^i)^C{}_B+G_{CB}(J^i)^C{}_A , &
		\check \nabla_A (J^i)^B{}_{C} &=0
		\label{gaugedNis4sum4}
	\end{tcolorbox}

	\subsection{Conditions for $D(2,1;\a)$ superconformal invariance}\label{SecD21aconds}
	We now turn to the conditions for the gauged $\caln=4B$  sigma model  to be conformally invariant. If these are obeyed we  will see that the full symmetry algebra belongs to the one-parameter family of superconformal algebras  $D(2,1;\a)$. These algebras contain a second $su(2)$ $R$-symmetry, and   the value of the parameter $\a$ will be determined by the transformation of the supercharges under this second set.
	
	As we did in the previous sections we first consider the off-shell realization of the symmetry algebra on the fields, independent of the invariance of the Lagrangian. Supersymmetry and gauge transformations require a quaternionic structure $J^i$ and Killing vectors $k_I$, and we have seen in Section \ref{Secbosconf} that conformal transformations are parametrized  in terms of a vector $\x$ and functions $h^I$. 
	To display the full algebra of transformations, it is useful to define, with the help of  (\ref{Jrhodef}),  vector fields $\o_\r^A$ and 
	one forms $V_A^{\r I}$ as follows
	\bea
	\omega_\r^A&=& (\bar J_\r)^A_{\ B} \xi_\perp^B, \qquad \r = 1,\ldots, 4  \label{defom}\\
	V_A^{\r I}&=& ( J^\r)^B_{\ A} \partial_B h^I.\label{defV} 
	\eea
	A rather lengthy analysis of the closure of the full symmetry algebra  yields the following conditions on the geometric data in addition to (\ref{gaugedbossum3},\ref{confbossum2},\ref{confbossum3},\ref{gaugedNis4sum1}-\ref{gaugedNis4sum4}). First of all, the vector field $\x^A$ should be triholomorphic:
	\be
	\call_\xi (J^i)^A{}_B =0.\label{xitrihol}
	\ee
	Furthermore, the composite vector fields $\omega_i^A$ in \eqref{defom} should transform the complex structures as
	\begin{equation}
		\call_{\omega^i} (J^j)^A{}_B=\frac{1}{1+\alpha}\epsilon_{ijk}(J^k)^A{}_B+k_I^A V^{i I}_C (J^j)^C{}_B-(J^j)^A{}_C k^C_I V^{i I}_B.\label{omJ}
	\end{equation}
	We note that this transformation law contains the parameter $\a$ which determines the algebra $D(2,1;\a)$ that is realized on the fields.
	A final condition is that the one-forms $V^{iI}_A$ should be closed:
	\begin{equation}
		\partial_{[A}V^{iI}_{B]}=0.
	\end{equation} 
	
	The resulting set of fermionic transformations can again be succinctly written in terms of
	an infinitesimal parameter $\Sigma_\r=\epsilon_\r +\eta_\r t$, where the time-independent Grassmann variables $\epsilon_\r, \eta_\r$ parametrize supersymmetry and superconformal transformations respectively. The fields transform as
	\begin{eqnarray}
		\delta_\Sigma x^A&=&-i(J^\rho)^A{}_B\Sigma_\rho \chi^B\label{deltaxD21a}\\
		\delta_\Sigma a^I&=&2i V_A^{\rho\, I}\dot \Sigma_\r \chi^A\label{afermmod}\\
		\delta_\Sigma \chi^A&=&(\bar J^\rho)^A{}_B\left(\Sigma_\rho D_t x^B+2\dot{\Sigma}_\r \xi_\perp^B\right)+i\partial_C (J^\r)^A{}_B\Sigma_\r \chi^C\chi^B\label{sigmachimod}
	\end{eqnarray}
	We note in particular that the gauge fields $a^I$ generically transform nontrivially under superconformal transformations.

	In addition, under commutators of  the fermionic transformations a second $su(2)$ $R$-symmetry  is  generated  in addition to (\ref{rtildegauged}). Parametrizing it by three parameters $r^i$ the fields transform as
	\begin{eqnarray}
		\delta_r x^A&=&(1+\alpha)\,r^i\omega_{i}^A\label{def omega}\nonu
		\delta_r a^I&=&-(1+\alpha)\,r^i V_A ^{i\, I}D_t x^A\label{ar1}\nonu
		\delta_r \chi^A&=&(1+\alpha)\,r^i\left(\partial_B\omega_i^{A}+V_B^{i I}k_I^A\right)\chi^B\label{rferm}
	\end{eqnarray}
	Our parametrization of two $su(2)$ actions is  natural and simple since (\ref{rferm})  acts geometrically while (\ref{rtildegauged})  acts only on the fermions. However, these two factors do  not commute. Two commuting $su(2)$ factors are generated by $\delta_{r_+}=\delta_{\tilde r}$, with $r_+^i=\tilde r^i$, and  $\delta_{r_-}=\delta_r + \delta_{\tilde r}$, with $r_-^i=r^i=\tilde r^i$. 
	
	The transformations (\ref{deltaxD21a}-\ref{rferm}) generate an off-shell realization of a gauged version of the $D(2,1;\a)$ algebra whose commutators are given   in  (\ref{D210sum7}-\ref{D210sum13}) below.
	One furthermore checks that the invariance of the Lagrangian under the  transformations above does not impose any  geometric conditions on the target space in addition to those required for $\caln = 4B$ supersymmetry and conformal invariance of the action\footnote{One can show that the  condition $\call_\xi C = - C$ follows from (\ref{xitrihol}).}. 
	Summarized,  gauged sigma models invariant under $D(2,1;\a)$  are constructed with the help of   the following structures in  addition to those listed in
	(\ref{gaugedbossum1}-\ref{gaugedbossum6},\ref{confbossum1}-\ref{confbossum6}, \ref{gaugedfermsum1}, \ref{conffermsum1}-\ref{conffermsum5}
	\ref{gaugedNis4sum1}-\ref{gaugedNis4sum4}):
	\begin{tcolorbox}[ams align]
		{\rm SCS: }& & \delta_\Sigma x^A&=-i(J^\rho)^A{}_B\Sigma_\rho \chi^B  &
		\delta_\Sigma a^I&=2i V_A^{\rho\, I}\dot \Sigma_\r \chi^A\label{D210sum1}\\ 
		&&	\delta_\Sigma \chi^A&=(\bar J^\rho)^A{}_B\left(\Sigma_\rho D_t x^B+2\dot{\Sigma}_\r \xi_\perp^B\right)  & &\hspace{-1.2cm}+i\partial_C (J^\r)^A{}_B\Sigma_\r \chi^C\chi^B && \\
		{\rm RS: }& &  \delta_r x^A&=\,(1+\alpha)r^i\omega_{i}^A &
		\delta_r a^I&=-(1+\alpha)\,r^i V_A ^{i\, I}D_t x^A\\
		&&	\delta_r \chi^A&=\,(1+\alpha)r^i\left(\partial_B\omega_i^{A}+V_B^{i I}k_I^A\right)\chi^B&&\\ {\rm \tilde RS: }&&\delta_{\tilde r}x^A&=0 & \delta_{\tilde r}a^I&=0\\
		&&	\delta_{\tilde r}\chi^A&=\frac{1}{2}\tilde r^i (J^i)^A{}_B\chi^B &&\\
		{\rm A:}&&  
		{}[\delta_{r_1},\delta_{r_2}]&=\delta_{r_3}+ \delta_\lambda &r_3^i&=\epsilon_{ijk}r^j_1r^k_2,\nonu
		&& &&\lambda^I&=(1+\a)^2\epsilon_{ijk}r^j_1r^k_2\call_{\omega_i}h^I\label{D210sum7}\\
		&&	{}[\delta_{\tilde r_1},\delta_{\tilde r_2}]&=\delta_{\tilde r_3} \, & \tilde r_3^i&=-\epsilon_{ijk}\tilde r^j_1\tilde r_2^k\\
		&&	{}[\delta_{r_1},\delta_{\tilde r_2}]&=\delta_{\tilde r_3} & \tilde r_3^i&=\epsilon_{ijk}r_1^j\tilde r_2^k\\
		&&	{}[\delta_{P},\delta_{\Sigma_1}]&=\delta_{\Sigma_2}& \Sigma_{2\rho}&=-P\dot \Sigma_{1\r}+\frac{1}{2}\dot P\Sigma_{1\r}\\
		&&	{}[\delta_{r},\delta_{\Sigma_1}]&=\delta_{\Sigma_2}+\delta_\lambda& \Sigma_{2\rho}&=\frac{1}{2}\left( j^i_- + j^i_+ \right)_{\rho \sigma} {r}^i \Sigma_{1\sigma}\nonu
		&& &&& \hspace{-0.5cm}\lambda^I = (1+\a)V^{i\, I}_A  r^i \delta_{\Sigma_1}x^A \\
		&&	{}[\delta_{\tilde r},\delta_{\Sigma_1}]&=\delta_{\Sigma_2}& \Sigma_{2\rho}&=- \frac{1}{2}(j_+^i)_{\r\s}\Sigma_\s \tilde r^i\\
		&&	{}[\delta_{\Sigma_1},\delta_{\Sigma_2}]&=\delta_{P}+\delta_{r}+\delta_{\tilde r}+\delta_\lambda &  P&=2i\Sigma_{1\r}\Sigma_{2\r}\nonu
		&& &&&\hspace{-4cm}	r^i=\frac{2i}{1+\alpha}(j^i_-)_{\r\s}(\dot\Sigma_{1\r}\Sigma_{2\s}-\Sigma_{1\r}\dot\Sigma_{2\s})\nonumber\\
		&& &&&\hspace{-4cm}\tilde r^i=\left(\frac{2\alpha i}{1+\alpha}j^i_+ -\frac{2i}{1+\alpha} j^i_-\right)_{\r\s}(\dot\Sigma_{1\r}\Sigma_{2\s}-\Sigma_{1\r}\dot\Sigma_{2\s})\nonumber\\
		&&&&&
		\hspace{-4.1cm}\lambda^I=-2i\left(a^I\Sigma_{1\r}\Sigma_{2\r}+h^I\frac{d}{dt}(\Sigma_{1\r}\Sigma_{2\r})\right)\label{D210sum13}\\
			{\rm SC:}&&  
	\call_\xi (J^i)^A{}_B &=0 &\partial_{[A}V^{iI}_{B]}&=0\label{D210sum14}\\
		&&&&&
	\hspace{-8.4cm}	\call_{\omega^i} (J^j)^A{}_B=\frac{1}{1+\alpha}\epsilon_{ijk}(J^k)^A{}_B+k_I^A V^{i I}_C (J^j)^C{}_B-(J^j)^A{}_C k^C_I V^{i I}_B\label{D210sum15}
	\end{tcolorbox}
	Here, the $j_\pm$ denote the (anti-)selfdual 't Hooft symbols given by
	\be 
	(j_\pm^i)_{\m\n} = \mp (\d_{\m i}\d_{\n 4} - \d_{\m 4}\d_{\n i} )- \e_{i\m\n{4}}.\label{jsdef}
	\ee

	\subsection{Quantization}
	We now address the issue of the operator ordering in the quantum generators of $D(2,1;\a)$, starting from  the canonical (anti-)commutators and Hermiticity properties of the fields given in (\ref{ccrferm}) and (\ref{Hermprop}). We  {find} that the quantized $D(2,1;\a)$ generators are given by
	\begin{tcolorbox}[ams align]
		Q^i &= \chi^A (J^i)_{\ \ A}^{B} \P_B + {i \over 2}  (J^i)_{[A}^{\ \ D} C_{BC]D} \chi^A \chi^B \chi^C\label{D210gen1}\\
		Q^4 &= \chi^A \P_A - {i \over 6} C_{ABC} \chi^A \chi^B \chi^C\label{D210gen2}\\
		S^\m &= - 2 \o^\m_A  \chi^A  \label{D210gen3}\\
		R^i &= (\a+1)\left( \o^{iA} \left(\P_A - {i\over 4} C_{ABC} \chi^B \chi^C\right)-{i\over 2}\nabla_A \o^{iA} -{i\over 2}\left( \nabla_{[A}\o^i_{B]}-k_{I[A} V^{i I}_{B]}\right) \chi^A \chi^B  \right) \\
		\tilde R^i & =  {i \over 4} J^i_{AB} \chi^A \chi^B,\label{D210gen4}
	\end{tcolorbox}
	where the expressions for $\P_A$  and the remaining generators $M_I', H', D', K$ are the same as in the $osp(1,2)$ case, see (\ref{ospgen1},\ref{ospgen4}-\ref{ospgen7}). Comparing these expressions with (\ref{sugen1}-\ref{sugen4}) we note  that $Q^\b, S^\b$ with $\b = 3,4$ generate an $su(1,1|2)$ subalgebra whose R-charge is the combination $R = - (1+\a)^{-1} R^3-  \tilde R^3$.
One furthermore  verifies the basic commutation relations involving the $R$-charges 
	\begin{align}
		 [ Q^4, \tilde R^i] & = {i \over 2} Q^i & 
		[ Q^i, \tilde R^j ] &= -{i\over 2} \left( \d^{ij} Q^4 - \e^{ijk} Q^k \right)&
	 \{ Q^4, S^i \}  &= -{2\over \a+1} R^i - 2\tilde R^i\label{D4}\\
	 	[ R^i,  R^j] &= - i \e^{ijk}  R^k & 
				[ R^i, \tilde R^j] &=-  i \e^{ijk} \tilde R^j &
		[\tilde R^i, \tilde R^j] &=  i \e^{ijk} \tilde R^k \label{D5}\\
		[M_I', \tilde R^i ] &=0 & 	[\tilde R^i, H' ] &=0 \label{D6}
	 	\end{align} 
 To prove the second identity in (\ref{D4}), one needs
	the following wHKT identity, proven in \cite{Smilga:2012wy},
	\bea
	\left(J^{i\ D}_A J^{j\ E}_B -J^{i\ D}_B J^{j\ E}_A \right)C_{CDE} + \left(J^{i\ D}_C J^{j\ E}_A -J^{i\ D}_A J^{j\ E}_C \right)C_{BDE}&\nonu  + \left(J^{i\ D}_B J^{j\ E}_C -J^{i\ D}_C J^{j\ E}_B \right)C_{ADE} = 2  \d^{ij} C_{ABC}.&
	\eea
The $R^i$ charges (\ref{D210gen4}) are obtained from verifying the last anticommutator in (\ref{D4}) and using the identity (\ref{omCid}).

The remaining	(gauged) $D(2,1;\a)$ commutation relations follow from the above  
elementary ones
upon applying the	 (super-)Jacobi identities. One obtains  for instance
	\begin{align}
	\{ Q^\m , Q^\n \} &= 2 \d^{\m\n} H' , \qquad 	\{ S^\m , S^\n \} = 2 \d^{\m\n} K \\
	\{ Q^\m, S^\n \} &= - 2 \d^{\m\n} D' +{2 \over 1+ \a} (j_-^i)_{\m\n} R^i - \left( {2 \a\over 1+ \a} (j_+^i)_{\m\n} - {2 \over 1+ \a} (j_-^i)_{\m\n} \right) \tilde R^i\\
	    	[Q^\m, \tilde R^i ] &={i\over 2} (j_+^i)_{\m\n} Q^\n, \qquad
	    		[S^\m, \tilde R^i ] ={i\over 2} (j_+^i)_{\m\n} S^\n 
			\end{align}
		For the remaining $D(2,1;\a)$  commutation relations we refer to e.g. \cite{Michelson:1999zf}.

	The generators once again simplify when expressed in complex coordinates $(z^m, \bar z^{\bar m})$ adapted to, say, the complex structure $J^3$, i.e
	\be 
	(J^3)_m{}^{n} = - i \d_m ^n, \qquad (J^3)_{\bar m}{}^{\bar n} =  i \d_{\bar m}^{\bar n}\label{complcoord}
	\ee 
	We define the complex super(conformal) charges
	\begin{align} 
		\calq_+ &= -\half (Q^1 + i Q^2), & \calq_- &= \half (-Q^3 + i Q^4)\\
		\cals_+ &=- \half (S^1 + i S^2), & \cals_- &= \half (-S^3 + i S^4).
	\end{align}
	These form doublets under the $\tilde R$-symmetry,
	\begin{align}
		[\calq_\pm, \tilde R_\mp] &= \calq_\mp, & [\calq_\pm, \tilde R_\pm] &=0 \\
		[\cals_\pm, \tilde R_\mp] &= \cals_\mp &[\cals_\pm, \tilde R_\pm] &= 0.
	\end{align}
	where
	\be
	\tilde R_\pm :=\tilde  R^1 \pm i \tilde  R^2.
	\ee
	The charges  $\calq_-, \cals_-$ take the form (\ref{Nis2gencc}) and the $\tilde R$-charges become
	\begin{align}
	\tilde R^3 &= {1\over 4} G_{m \bar n} ( \chi^m \chi^{\bar n} -\chi^{\bar n}\chi^m ), \qquad 
	\tilde R_+ = {i \over 2} (J_+)_{mn}  \chi^m \chi^{ n}, &
	\tilde R_- &= {i \over 2} (J_-)_{\bar m \bar n}  \chi^{\bar m} \chi^{\bar n},\label{Nis4gencc2}
	\end{align}
where we have used that $(J^+)_{AB}$ is of type (2,0) with respect to the complex structure $J^3$.

	The algebra can  once more be realized on the Hilbert space of antiholomorphic forms as worked out in Section \ref{Secdiffform} above. The expressions for the  (conformal) supercharges $\calq_-,\cals_-$ were already given in  (\ref{Nis2diff0}-\ref{Nis2diff2}). 	We remark  that, when the background gauge field  $A_A=0$ vanishes, our expression (\ref{Nis2diff1}) for $\calq_-$  coincides with the one given in \cite{Smilga:2012wy}, where the supercharges were derived following a  different route. 
Furthermore, the gauge generators and $\tilde R$-charges are  realized as 
	\begin{tcolorbox}[ams align]
		\tilde R^3  &=  {d \over 8} - \half ({\rm form\ degree}), \qquad 
		\tilde R_-  = i J_- \wedge , \qquad
		\tilde R_+  = {i\over 2} (J_+)_{\underline m \underline n } {\d \over \d e^{\bar{\underline m}}}
		{\d \over \d e^{\bar{\underline n}}},\label{Nis4diff2}\\
			M_I'
			  &=  -i \call_{k_I} - (i_{k_I} A + v_I) -{i \over 2} f_{IJ}^{\ \ J}+ {i \over 2} k^A \check \o_{A n \bar m} G^{n \bar m},  
		\label{Nis4diff1}
	\end{tcolorbox}
	where $\check \o_{A B C}= \o_{A B C} - \half C_{ABC}$ is the torsionful spin connection.
	The $\tilde R$-generators realize a standard Lefschetz-like $su(2)$-action on the space of antiholomorphic forms on quaternionic  manifolds  \cite{Moore:2015szp,verbitsky2002hyperkaehler}. 
The supersymmetric ground states  correspond  to elements of the cohomology  of the twisted Dolbeault operator (\ref{Nis2diff1}) which are
in addition 
annihilated by $\tilde R_+$. 

The expression for the gauge charges $M_I'$ arises from (\ref{MIcc}) upon using tri-holomorphicity of the Killing vectors $k_I$, which in particular leads to 
\be
\left( \pa_{[\bar n} k_{p]}+\half k^A C_{\bar n p A}\right) G^{\bar n p} =0.
\ee

\section{Examples with $D(2,1;\a)$ superconformal symmetry} 
\label{Sec341}	
	To illustrate the general discussion of the previous sections, we  give an explicit class of nontrivial examples of our construction. In these,  $\caln=4$ supersymmetric  sigma models acquire $D(2,1;\a)$ superconformal symmetry upon gauging an isometry group. These examples arise from recasting $D(2,1;\a)$ models formulated in terms of 
	$\textbf{(3,4,1)}$ multiplets (each containing with three real bosons, 4 real fermions and one auxiliary field) into a more geometric  $\textbf{(4,4,0)}$ multiplet description. These models are also physically interesting, as they include as a special subclass the Coulomb branch dynamics of multi-centered D-brane systems which develop an AdS$_2$ throat region. Besides being relevant for the (n)AdS$_2$/(n)CFT$_1$ correspondence these provide an infinite set of examples of our construction possessing $D(2,1;0)$ symmetry and with a fully explicit Lagrangian.
	
	\subsection{Gauged $\caln = 4$ sigma models with a $\textbf{(3,4,1)}$ origin}
$\caln=4$ supersymmetric  sigma models  which are formulated in terms of $n$ 
	$\textbf{(3,4,1)}$ multiplets can be recast as {\em gauged} sigma models with $n$ $\textbf{(4,4,0)}$ multiplets \cite{Ivanov:2011gk}, see \cite{Mirfendereski:2020rrk} for a review in the component formulation used in the current work.
	In  models obtained in this way, the target space coordinates $x^A$ are regrouped into quadruples $x^{ \m a}$, where $\m = 1, \ldots 4$ and $a = 1, \ldots, n$.  The coordinates $x^{i a}, i = 1,2,3$  come from the dynamical bosons in the $\textbf{(3,4,1)}$ language, and the $x^{4 a}$ originate from the auxiliary fields. The
	Lagrangian is of the form (\ref{gaugedbossum1},\ref{gaugedfermsum1}) where the target space gauge field $A_{\m a}$, metric $G_{\m a\, \n b}$, torsion tensor $C_{\m a\, \n b\, \r c}$ and complex structures $(J^i)^{\m a}{}_{\n b}$ are  independent of the $x^{4a}$ and of the form\footnote{The condition (\ref{341conds2}) is not strictly necessary in the 	$\textbf{(3,4,1)}$ formulation, though it is unclear whether a $\textbf{(4,4,0)}$ reformulation exists if it is not obeyed, see \cite{Mirfendereski:2020rrk}.}
	\bea
	G_{\m a\, \n b}&=&\delta_{\m\n}G_{ab}\\
	\partial_{ia}G_{bc}&=&\partial_{i(a}G_{bc)}\label{341conds2}\\
C_{\m a\, \n b\, \r c}&=&\partial_{\l a}G_{bc}\,\epsilon_{\l\m\n\r}\label{CitoG}\\
(J^i)^{\m a}{}_{\n b}&=& (j^i_+)_{\m\n}\delta^a_b\label{j+}\\
\pa_{4 a} A_{\m b} &=& \pa_{4 a} G_{bc} =0,
	\eea 
	with $(j^i_+)_{\m\n}$ defined in (\ref{jsdef}). Supersymmetry furthermore requires that the background gauge field strength is self-dual and that the metric $G_{ab}$ is determined by an $x^{4a}$-independent potential function $\calh$:
	\bea
	F_{\m a\, \n b}&=&	-\frac{1}{2}\epsilon_{\m\n\r\s}F_{\r a\, \s b}\\
	G_{ab} &=&\half\pa_{ia}\pa_{ib}  \calh.\label{GitoH}
	\eea
Using these relations one can  show that the requirements for $\caln = 4$ supersymmetry (\ref{Nis4ung1}-\ref{Nis4ung3}) are satisfied \cite{Mirfendereski:2020rrk}.	
	
From the above it is clear that the target space possesses $n$ commuting  Killing vectors
\be 
k_a = \pa_{4a}, \qquad a = 1, \ldots , n.
\ee
We will consider sigma models where the corresponding symmetry is gauged, so that the gauge algebra is $\mathfrak{g}= u(1)^n$ in this case. 
 Classically, the gauging imposes a symplectic reduction under the constraints
\be
M_{a} \approx 0
\ee
and is required for the model to posses an equivalent $\mathbf{(3,4,1)}$ description.
The potentials $v_a$ defined through (\ref{defvs}) are
\be 
v_a = - A_{4a}.
\ee

\subsection{$D(2,1; \a)$-invariant subclass}
In the covariant approach adopted in this work, conformal symmetry is parametrized by a target space vector  $\x$ which in these examples takes the form 
\be 
\x = \g x^{  \m a} \pa_{\m a},
\ee
where the constant $\g$ will determine the parameter $\a$ in the $D(2,1; \a)$ symmetry algebra. The derived objects $h^a, \g^a_{\ b}, \x_\perp, \o_i, V^i$ defined in Section \ref{SecD21aconds} are found to be
\begin{align}
h^a &= \g x^{4 a}, & \x_\perp &= \g x^{ia} \pa_{i a}, & \g^a_{\ b} &= \g \d^a_b\label{derquants1} \\
\omega_i&=-\g\epsilon^{ijk}x^{ja}\partial_{ka}-\g x^{ia}\partial_{4a}\ , & V^{ia}&=\g dx^{ia}. & 
\end{align}
One checks that these satisfy the structural constraints (\ref{D210sum13}-\ref{D210sum14}) required to have a closed $D(2,1;\a)$ algebra. In particular, verifying (\ref{D210sum15}) one finds that the parameters $\a $ and $\g$ are related as 
\be
\g=\frac{1}{1+\a}.\label{gitoa}
\ee

The conditions (\ref{confbossum5},\ref{confbossum7}, \ref{conffermsum5})  for the action to be $D(2,1;\a)$-invariant reduce to
\begin{align}
x^{jb}\partial_{jb}A_{4a}&=-A_{4a}, &
x^{jb}\partial_{jb}A_{ia}&=x^{jb}\partial_{ia}A_{jb} &&\label{exAcond}\\
 x^{ic}\partial_{ic}G_{ab}&=-(\gamma^{-1}+2)G_{ab}, & G_{ab}x^{ib}&=-\gamma\partial_{ia}(G_{bc}x^{jb}x^{jc}), &
\epsilon_{ijk}x^{ja}\partial_{ka}G_{bc}&=0\label{exmetrcond}
\end{align}
The last requirement, which originates from the condition (\ref{conffermsum5}) on the torsion tensor, imposes a rotational invariance condition on the metric. 

Since the conformal Killing vector $\x$ is not orthogonal to the Killing vectors $k_a$ (or, equivalently, the functions $h^a$ in (\ref{derquants1})  are nontrivial), if the above conditions are met we obtain  nontrivial examples of our construction, where $D(2,1;\a)$ invariance only appears after gauging, i.e. after reduction to the constraint surface $M_a \approx 0$. 

	\subsection{Explicit examples}
	We now give two sets of explicit examples. The simplest class arises from metrics of the form
	\begin{equation}
	G_{ab}=r^{-(\alpha+3)} \delta_{ab}\qquad r^2=x^{ia}x^{ia}
	\end{equation}
	One verifies that indeed all of \eqref{exmetrcond} are satisfied with $\g = (1 + \a)^{-1}$ as required for $D(2,1;\a)$ invariance\footnote{One could extend these models to include a background electromagnetic field of the type discussed in the second class of examples.},  
	see  \eqref{gitoa}. 
	
	The second class of examples arises in the description of $n$-centered D-brane systems in a certain scaling limit of the charges, where an AdS$_2$ throat forms. In a suitable regime these allow for an effective description as an $n$-node Coulomb branch quiver quantum mechanics \cite{Anninos:2013nra},\cite{Mirfendereski:2020rrk}. When reformulated in the $\textbf{(4,4,0)}$ language these give rise  to models of the type considered here. In particular, the background electromagnetic fields $A_{4a}$ and $A_{ia}$ take the form of the Coulomb potential and magnetic monopole vector potential felt by the $a$-th center due to the  other centers:
	\be
	 A_{4a}= -\sum_{b, b\neq a} \frac{\kappa_{ab}}{2r_{ab}}, \qquad A_{ia}=- \sum_{b,b\neq {a}} \kappa_{ab}\,A_i^{\mathrm{D}}(\tilde x_{ab}),
	\ee
	 where $A_i^D$ is the Dirac monopole vector potential
	 \be
	 \qquad A_i^{\mathrm{D}}(x)=\frac{\epsilon_{ijk} n^j x^k}{2 \,r(x^ln^l-r)},
	 \ee 
with $n^i$ is an arbitrary  unit vector indicating the direction of the Dirac string), and we   defined
	\begin{equation}
	\tilde x_{ab}=\begin{cases}
	x_{ab}=x_a-x_b&\quad\mbox{when } a<b\\
	x_{ba}=x_b-x_a&\quad\mbox{when } a>b.
	\end{cases}\label{tildex}.
	\end{equation}
	One verifies that these obey the conditions (\ref{exAcond}) for conformal invariance. 
	The microscopic origin \cite{Denef:2002ru} of the parameters $\k_{ab}= - \k_{ba}$  is as  Dirac-Schwinger-Zwanziger  inner products of the D-brane charges of the   centers labelled by $a$ and $b$.
	
	The potential function $\calh$ from which the metric and torsion are derived through (\ref{GitoH},\ref{CitoG}) is  given by
\be	\calh = -\sum_{a,b, b\neq a} \frac{|\kappa_{ab}|}{4r_{ab}}\log r_{ab}.
\ee
This second class of examples obeys the conformal invariance conditions  (\ref{exmetrcond}) for $\g=1$, so that the symmetry algebra of these models is $D(2,1;0)$, which is isomorphic to the semi-direct product $\mathfrak{psu(1,1|2)}\rtimes\mathfrak{su}(2)$. We refer the reader to \cite{Mirfendereski:2020rrk} for a detailed discussion of the symmetries of these models in their $\textbf{(3,4,1)}$ description.
Let us also mention that the above quiver models describe the relative motion of the D-branes and are subject to a further constraint imposing the decoupling of the center-of-mass degree of freedom, see again \cite{Mirfendereski:2020rrk} for details.
	
\section{Outlook}
In this work we undertook a systematic study of the constraints imposed by (super) conformal invariance on the  geometry of the target space $\calm$ in gauged sigma models. Our main conclusion was that apart from models that are conformal with or without gauging of isometries, there are also models that only are conformally invariant when gauged. For models of the second type the conditions on the geometry of $\calm$ are a  deformation of those encountered in ungauged models. This is in particular relevant in the supersymmetric case, where $\calm$ exhibits a torsional (hyper) Kahler geometry, which is absent in the formulation of the model on $\calm/\mathrm{G}$, as obtained after integrating out the gauge fields.

In the supersymmetric setting we focused on type B models, where  the quantum symmetry algebra has an explicit  realization on the Hilbert space of antiholomorphic differential forms. 
As we illustrated in section \ref{Sec341}, the formulation of some models in this geometric language  necessarily requires some symmetries to be gauged. 

An application of, and motivation for, the present work is provided  by the Coulomb branch quiver mechanics describing   the dynamics  D-brane systems in an AdS$_2$ scaling limit. These systems are  important    due to their connection to (n)AdS$_2$/(n)CFT$_1$ and black hole physics. A first step \cite{jorisetal}  would be to use our formulation to study the  explicit Hilbert space and it's $D(2,1; \a)$ content  in the simplest examples discussed in  Section \ref{Sec341}.

More generally, it would be of great interest to learn more about the quantum Hilbert space through an appropriate superconformal index (see \cite{Gaiotto:2004pc} for the $D(2,1;\a)$ algebra)  and to develop  localization methods for its computation.  An especially interesting question in our view is whether the model contains any singlet ground states under the full superconformal symmetry, which would be candidate microstates dual pure  AdS$_2$  in the picture of \cite{Sen:2008yk}. From our representation of the conformal generators it is clear that   singlet states  can exist  only under very special circumstances. Indeed, the special 
conformal generator $K = 2 (\x_\perp)^2$ can only annihilate a state if it's wavefunction  has support on the fixed locus of the vector $\x_\perp$. This is essentially what happens in the $\caln = 4$ `type A'   models with hyperk\"{a}hler target spaces \cite{Dorey:2018klg},\cite{Dorey:2019kaf}, which were shown to contain conformal singlet  ground states. In these models, the $\x_\perp$ fixed locus is singular and the result requires a  supersymmetry-preserving resolution of this singularity.  It  would be interesting to see if a similar mechanism can occur in the type B models considered in this work. 

Finally let us remark that it it would be interesting to investigate a possible role of the 'radial-angular' split of (super)conformal mechanics, see e.g. \cite{Hakobyan:2008xx,Hakobyan:2009ac,Khastyan:2021qlx} , in the considerations above.


	\section*{Acknowledgements}
	DVdB and DM were partially supported by TUBITAK grant 117F376, in addition DVdB was partially supported by the Bilim Akademisi through a BAGEP award. C\c{S} was partially supported by a TUBITAK 2214-A Research Fellowship Programme for PhD Students. The research 	of JR was supported by the  European Structural and Investment Fund and the Czech Ministry of Education, Youth and Sports (Project CoGraDS - \linebreak CZ.02.1.01/0.0/0.0/15\textunderscore 003/0000437).
	
	\appendix

	\section{Useful identities}
	Here we list a number of useful identities one can further verify once the geometric objects $ (J^i,\xi^A,k^A_I,h^I) $ satisfy the set of structural conditions for $ D(2,1;\a) $ superconformal algebra given in (\ref{gaugedbossum3},\ref{confbossum2},\ref{confbossum2b},\ref{gaugedNis4sum1},\ref{gaugedNis4sum2},\ref{D210sum14},\ref{D210sum15}). For more convenience we repeat the following definitions
	\begin{equation}
		\xi_\perp^A=\xi^A-h^Ik_I^A\ , \qquad \omega_i^A= -(J_i)^A_{\ B} \xi_\perp^B\ , \qquad V_A^{i I}= ( J^i)^B_{\ A} \partial_B h^I\ . 
	\end{equation}
	One then checks	
	\begin{eqnarray}
		\mathcal{L}_{\xi}\xi_\perp^A&=&0\\ [2pt]
		\mathcal{L}_\xi \omega_i^A&=&0\\ [2pt]
		\mathcal{L}_{k_I}\xi_\perp^A &=&0\\ [2pt]
		\mathcal{L}_{\omega_i}\xi_\perp^A&=&k^A_I V^{iI}_B \xi^B_\perp\\ [2pt]
		\mathcal{L}_{\omega_i}k_I^A&=&0\\ [2pt]
		\mathcal{L}_{\omega_i}\omega_j^A&=&\epsilon_{ijk}\Big(\frac{1}{1+\a}\o^A_k+k^A_I\mathcal{L}_{\o_k}h^I\Big) \\ [2pt]
		\mathcal{L}_{\omega_i}V^{jI}_A&=&\frac{1}{1+\a}\epsilon_{ijk}V^{kI}_A-2V^{J[i}_A V^{j]I}_B k_J^B\\ [4pt]
		\mathcal{L}_\xi V^{iI}_A&=&\gamma^I{}_J V^{iJ}_A\\ [2pt]
		\mathcal{L}_{\xi_\perp} V^{iI}_A&=&k_J^B\big(\partial_B h^I V^{iJ}_A-\partial_A h^J V_B^{iI}\big) \\ [2pt]
		\mathcal{L}_{k_J}V^{i I}&=&-f_{JK}{}^I V^{iK}_A\\ [2pt]
		\mathcal{L}_{\xi_\perp}h^I&=&0\\ [2pt]
		\mathcal{L}_{\omega_i}h^I&=&-V_{A}^{iI}\xi_\perp^A\\
			\o^{iA}C_{ABC} &=& J^i_{BC} + 2 \nabla_{[B} \o^i_{C]} + 2 k_{I[B} V^{i I}_{C]}\label{omCid}.
	\end{eqnarray}

\providecommand{\href}[2]{#2}\begingroup\raggedright\endgroup

\end{document}